\documentclass[pdflatex,sn-mathphys-ay]{sn-jnl}

\usepackage{graphicx}%
\graphicspath{{images/}} 

\usepackage{multirow}%
\usepackage{amsmath,amssymb,amsfonts}%
\usepackage{amsthm}%
\usepackage{mathrsfs}%
\usepackage[title]{appendix}%
\usepackage{xcolor}%
\usepackage{textcomp}%
\usepackage{manyfoot}%
\usepackage{booktabs}%
\usepackage{algorithm}%
\usepackage{algorithmicx}%
\usepackage{algpseudocode}%
\usepackage{listings}%

\usepackage{subcaption}
\usepackage{gensymb} 

\theoremstyle{thmstyleone}%
\newcommand{\gammadot}{\dot{\gamma}}

\usepackage{physics}
\usepackage[normalem]{ulem}

\newcommand{\vc}[1]{\boldsymbol{#1}}
\newcommand{\vt}[1]{\boldsymbol{\mathsf{#1}}}
\newcommand{\vv}[1]{\mathsf{\boldsymbol{#1}}}

\begin{document}

\title[Integral-fractional K-BKZ SPH LAOS]{Integral fractional viscoelastic models in SPH: LAOS simulations versus experimental data}


\author*[1]{\fnm{Luca} \sur{Santelli}}\email{lsantelli@bcamath.org}

\author[2]{\fnm{Adolfo} \sur{Vázquez-Quesada}}\email{a.vazquez-quesada@fisfun.uned.es}
\author[3]{\fnm{Aizeti} \sur{Burgoa}}\email{aburgoa@leartiker.com}
\author[3]{\fnm{Aitor} \sur{Arriaga}}\email{aarriaga@leartiker.com}
\author[3]{\fnm{Rikardo} \sur{Hernandez}}\email{rhernandez@leartiker.com}

\author[1,4,5]{\fnm{Marco} \sur{Ellero}}\email{mellero@bcamath.org}

\affil*[1]{\orgdiv{} \orgname{Basque Center for Applied Mathematics}, \orgaddress{\street{Mazarredo Zumarkalea, 14}, \city{Bilbao}, \postcode{48009}, \state{Bizkaia}, \country{Spain}}}

\affil[2]{\orgdiv{Facultad de Ciencias}, \orgname{Universidad Nacional de Educación a Distancia}, \orgaddress{\street{ C. de Bravo Murillo, 38}, \city{Madrid}, \postcode{28015}, \country{Spain}}}

\affil[3]{Sustainable Transport group, Polymers Technology, Leartiker S. Coop, \city{Markina-Xemein}, \postcode{48270}, \country{Bizkaia (Spain)}}

\affil[4]{\orgdiv{IKERBASQUE}, \orgname{Basque Foundation for Science}, \orgaddress{\street{Calle de María Díaz de Haro 3}, \city{Bilbao}, \postcode{48013}, \country{Spain}}}

\affil[5]{\orgdiv{Complex Fluid Group, Department of Chemical Engineering}, \orgname{Swansea University}, \orgaddress{\street{SA1 8EN}, \city{Swansea}, \country{United Kingdom}}}


\abstract{
The rheological behaviour of a polymer was investigated by performing numerical simulations in complex flow and comparing them to experiments. For our simulations, we employed a Smoothed Particle Hydrodynamics scheme, utilising an integral fractional model based on the K-BKZ framework. The results are compared with experiments performed on melt-state isotactic polypropylene under medium and large amplitude oscillatory shear. 

The numerical results are in good agreement with the experimental data, and the model is able to capture and predict both the linear and the non-linear viscoelastic behaviours of the polymer melt. Results show that equipping SPH with an integral fractional model is a promising approach for the simulation of complex polymeric materials under realistic conditions.
}




\maketitle

\section{Introduction}\label{sec:introduction}
Large Amplitude Oscillatory Shear (LAOS) is a powerful rheological technique used to investigate the nonlinear viscoelastic properties of complex fluids, such as polymer melts, colloidal suspensions, and biological materials. Unlike Small Amplitude Oscillatory Shear (SAOS), which probes the linear viscoelastic regime, LAOS applies larger deformations to the material, allowing researchers to explore the material's response under conditions that are more representative of real-world applications \citep{hyun2011review,yosick1997kinetic}. The technique provides detailed insights into the material's structure and dynamics by analysing the stress response as a function of strain amplitude and frequency. This information is crucial for understanding the material's behaviour under more realistic processing and end-use conditions, enabling the development of better-performing materials and products. LAOS experiments can reveal phenomena such as strain hardening, strain softening, thixotropy and yielding, which are not accessible through linear viscoelastic measurements \citep{blackwell2014simple,koekuti2016exploring,vargas2023multiscale}. Consequently, LAOS has become an essential tool in the field of rheology for characterising the complex mechanical behaviour of advanced materials.

A proper description of LAOS requires a rheological model that is able to capture the intrinsic non-linear behaviour of the complex fluid when subjected to large deformations \citep{hyun2011review,vargas2023multiscale}. A number of these descriptions relies on the framework provided by the integral Kaye–Bernstein, Kearsley, Zapas (K-BKZ) model \citep{bird1987dynamics,owens2002computational}. The model is based on the idea that the stress response of a material can be expressed as a convolution of the material's relaxation modulus with the strain history. Indeed, it has been shown that a polymer can be naturally described by employing a generalised integral-fractional K-BKZ constitutive model that captures both the power-law behaviour, typical of polymeric materials, and the non-linear response induced by the flow \citep{jaishankar2014fractional,santelli2024smoothed}. Fractional models describe the rheological behaviour of a fluid by introducing a constitutive element known as the spring-pot, which interpolates the elastic response of a spring and the viscous response of a dashpot \citep{gemant1938fractional,blair1942classification,jaishankar2013power,jaishankar2014fractional,yang2010constitutive,santelli2024smoothed}. The mathematical representation of the spring-pot is given by the fractional derivative, which is a generalization of the classical derivative that allows for non-integer order derivatives. 
However, solving the K-BKZ equation analytically (with or without the fractional formulation) can be challenging, due to the non-linearities and the complex conditions that arise in LAOS experiments. As a result, numerical simulations have become an essential tool for studying the rheological behaviour of complex fluids under LAOS.
Notably, the fractional derivative can be solved by an integral equation, naturally suggesting the adoption of an integral Lagrangian model for the numerical scheme \citep{larson1999structure,santelli2024smoothed}.

While integral models present several advantages, such as the ability to capture the long-term memory effects of complex fluids, they also come with challenges. Indeed, the integral equation can be computationally expensive and time-consuming, especially due to the need of tracking particle position \citep{viriyayuthakorn1980finite,rasmussen1993simulation,keunings2003finite,owens2002computational}.
A possible way to approach this issue is to perform simulations using a numerical method written in a Lagrangian formulation, thus overcoming the particle history tracking problem. We have shown that the Smoothed Particle Hydrodynamics method (SPH) is well equipped for this task, as it is a Lagrangian particle-based method that can handle large deformations and complex geometries, including constitutive equations written in integral fractional form \citep{ellero2002viscoelastic,ellero2005sph,vazquezquesada2009smoothed, santelli2024smoothed}, while experiencing only a minor loss in performance and accuracy.


In this article, we investigate the rheological behaviour of a polymer melt under LAOS using numerical simulations based on the Smoothed Particle Hydrodynamics method. This manuscript is organised as follows: 
in Section 2, we discuss the theoretical framework behind the integral approach and LAOS rheology; in Section 3, we describe the experimental setup and the numerical model used in this study; in Section 4, we present the results of the simulations and compare them to experimental data; and in Section 5, we discuss the implications of our findings and suggest future research directions.

\section{Theoretical integral framework}\label{sec:theory}
The integral representation of the constitutive equation can be understood as a simple mathematical transformation over the more common differential approach; however, its mathematical formulation is not-trivial. We report here our implementation, which strictly mimics the construction of \cite{bird1987dynamics}.

The main variable of this analysis is the displacement function $\vb{r} = \vb{r}(\vb r', t', t)$, which tracks the current position $\boldsymbol{r}$ of a particle, at time $t$, as a function of its past position $\vb r'$ at time $t'$. In this representation, the pair $\vb r', t'$ identifies the particle we are tracking. The Cartesian components of $\boldsymbol{r}$ are identified by $x_\mu$, with the Greek index ranging between the number of dimensions.
The velocity field can be directly derived by the displacements:
$\vb v'(\vb r',t')= \pdv{\vb r'(\vb r,t,t')}{t'}$ and $\vb v(\vb r,t)= \pdv{\vb r(\vb r',t',t)}{t}$.

From the displacement function we define a displacement gradient
\begin{equation}
    \Delta_{\mu\nu} (\vb r , t, t') = \pdv{x'_\mu(\vb r,t,t')}{x_\nu}
    \label{eq:grad_r'_bird}
\end{equation}
i.e. the gradient of the past position of the fluid particles at time $t'$ around the current position $\vb r$.
The equivalent form in terms of displacement of the current position around the previous is 
\begin{equation}
    E_{\mu\nu} (\vb r , t, t') = \pdv{x_\mu(\vb r',t',t)}{x'_\nu}.
\end{equation}
Together, they satisfy $\sum_{\mu,\nu}\Delta_{\mu\nu}E_{\mu\nu}=\delta_{\mu\nu}$ and, for incompressible fluids, $\det(\Delta_{\mu\nu})=\det(E_{\mu\nu})=1$.

From $\vb \Delta$ and $\vb E$, we construct, respectively, the Cauchy ($\vt B^{-1} $) and Finger ($\vt B$) finite strain tensors, the components of which are defined as follows:
\begin{align}
&B^{-1}_{\mu\nu}(\vb r,t,t') = \sum_\xi\pdv{x'_\xi}{x_\mu}\pdv{x'_\xi}{x_\nu}\\
&B_{\mu\nu}(\vb r,t,t') = \sum_\xi\pdv{x_\xi}{x'_\mu}\pdv{x_\xi}{x'_\nu}.
\end{align}
These are symmetric positive definite tensors, which are invariant under rigid rotation and translation. Thus they can be used to build finite strain tensors that vanish for rigid body motions:
\begin{align}
    &\boldsymbol{\gamma}^{[0]} = \vt B^{-1}-\vt 1\\
    &\boldsymbol{\gamma}_{[0]} = \vt 1 - \vt B
\end{align}
where $\vt 1$ is the identity tensor and $\boldsymbol{\gamma}^{[0]}$ and $\boldsymbol{\gamma}_{[0]}$ are the so-called relative finite strain tensors. 
We note that, while for small displacements $\boldsymbol{\gamma}^{[0]}=\boldsymbol{\gamma}_{[0]}$, this is, in general, not true for large displacements. Comparison with experimental data shows that more realistic constitutive models, e.g. the corresponding integral version of the upper convected Maxwell,  are built upon the finite strain tensor $\boldsymbol{\gamma}_{[0]}$.

The principal invariants of the Finger strain tensor $\vt B$, defined as the coefficients of its characteristic polynomial, are 
\begin{equation}
  I_1=\tr\vt B=\tr(\vt 1-\vt\gamma_{[0]}),
\end{equation}
\begin{equation}
  I_2=\tfrac12[(\tr\vt B)^2-\tr\vt B^2]=\tr\vt B^{-1}=\tr(\vt 1+\vt\gamma^{[0]}),
\end{equation}
and $I_3=\det\vt B$, which, for an incompressible fluid, equals one.
These scalars are used to construct a frame-indifferent model.

\subsection{Fractional K--BKZ model}
For the constitutive prescription of the incompressible viscoelastic fluid, we adopt the formulation of the fractional K--BKZ model. The classical K--BKZ model assumes the existence of a function $V$, scalar in its arguments $t-t',I_1,I_2$, and takes the nonlinear single-integral equation for the stress tensor of the form
\begin{equation}
    \vt \tau(t) = \int_{-\infty}^t
    \left[
    \frac{\partial V(t-t',I_1,I_2)}{\partial I_1}\vt\gamma_{[0]}(t,t')+
    \frac{\partial V(t-t',I_1,I_2)}{\partial I_2}\vt\gamma^{[0]}(t,t')
    \right]
    \dd t'.
    \label{eq:K--BKZ}
\end{equation}
It is customarily assumed separability of $V$ in time-dependent and strain-dependent factors:
\begin{equation}
    V(t-t',I_1,I_2) = M(t-t')U(I_1,I_2)
    \label{eq:}
\end{equation}
with $M$ the so-called \emph{memory} function and $U$ the elastic potential function, with the consistency condition with the linear limit $\left(\frac{\partial U}{\partial I_1}\right)_{3,3}+\left(\frac{\partial U}{\partial I_2}\right)_{3,3}=1$. To guarantee well-posedness for finite positive $M$, it is also required that $U$ is monotone in each argument, strictly monotone in $I_1$ or $I_2$, and convex in $\sqrt{I_1}$ and $\sqrt{I_2}$. The memory function $M$ is related to the so-called \emph{relaxation modulus} $G$ via the relation 
\begin{equation}
    M(t-t')=\frac{\partial G(t-t')}{\partial t'}.
\end{equation}
The factorized K--BKZ equation reads
\begin{equation}
    \vt \tau(t) = \int_{-\infty}^tM(t-t')
    \left[
    \frac{\partial U(I_1,I_2)}{\partial I_1}\vt\gamma_{[0]}(t,t')+
    \frac{\partial U(I_1,I_2)}{\partial I_2}\vt\gamma^{[0]}(t,t')
    \right]
    \dd t'.
    \label{eq:K--BKZ_factorized}
\end{equation}
In the following, we neglect the dependence of $U$ on $I_2$, so that the second term in the integral vanishes. This is consistent with the previous claim of choosing $\gamma_{[0]}$ as the preferred option for the finite strain tensor. 

The model that we adopt is commonly known as a fractional implementation of the Wagner integral model, originally proposed by \cite{wagner1976analysis} and later extended to the fractional case in several recent works, including \cite{jaishankar2014fractional, rathinaraj2022cox}. 

\subsubsection{Shear deformation}
We consider the unsteady homogeneous simple shear flow for a fluid confined between two parallel planes. We write $\tau$ for the shear component $\tau_{yx}$ of the stress tensor and $\gamma$ for the shear component $\gamma_{yx}$ of the strain tensor $\vt{\gamma}_{[0]}$. In this regime, the components of the velocity of the fluid are then $v_x=\dot{\gamma}(t)y$, $v_y=0$, $v_z=0$. We introduce a function $h$, called the \emph{damping} function, defined by $h(\gamma)=\tfrac{\partial U(I_1)}{\partial I_1}$, where we used the fact that, for this flow, $I_1=3+\gamma^2$. 
The equation for the shear component of the stress can then be written as
\begin{equation}
    \vt \tau(t) = \int_{-\infty}^tM(t-t')
    h(\gamma)\vt\gamma(t,t')
    \dd t'
\end{equation} 
where 
\[
\gamma(t,t')=\int_t^{t'}\dot{\gamma}(t'')\dd t''=\gamma(t')-\gamma(t).
\]

We take a generalised form of the damping function known as the Soskey-Winter damping function \citep{soskey1984large}:
\begin{equation}
    h(\gamma)=\frac{1}{1+(\gamma/\gamma^*)^m}
    \label{eq:damping_function}
\end{equation}

with $\gamma^*, m$ fitting parameters, which need to be determined experimentally. The physical meaning of these parameters is the following: the critical strain parameter $\gamma^*$ is related to the onset of nonlinearity (e.g. a lower value of $\gamma^*$ means that the non-linear effects become relevant at a lower strain); the damping exponent $m>0$ controls the rate at which the damping function decreases with increasing strain, and for $m=2$ one recovers the Doi-Edwards constitutive model for entangled polymers \citep{rathinaraj2021incorporating,rathinaraj2022cox,das2024laun}.

We adopt the Fractional Maxwell Model (FMM) for the shear stress-strain relation, which consists of two spring-pots, each characterised by a pair of material parameters, denoted by $\mathbb V$, $\alpha$ and $\mathbb G$, $\beta$, respectively, arranged in series \citep{jaishankar2014fractional}. These parameters are often written in terms of a characteristic time $\lambda_c = (\mathbb V / \mathbb G )^{1/(\alpha-\beta)}$ and a characteristic modulus $G_c=\mathbb V \lambda_c^{-\alpha}$. Namely, the FMM equation reads
\begin{equation}
    \tau(t) + \frac{\mathbb V}{\mathbb G}\dv[\alpha-\beta]{\tau(t)}{t}=\mathbb V \dv[\alpha]{\gamma(t)}{t}
\end{equation}
where the Caputo fractional derivative of order $\beta$ in time is defined as
 \begin{equation}
    \dv[\beta]{\vt\gamma(t)}{t} \equiv \frac{1}{\Gamma(1-\beta)}\int_0^t (t-t')^{-\beta}\dot{\vt\gamma}(t')\dd t' 
 \end{equation}
 with  $\Gamma(\cdot)$ the complete Gamma function and $0\le\beta\le 1$. 

The relaxation modulus and memory function associated with the Fractional Maxwell Model are then
\begin{align}
    &G(t-t') = \mathbb G (t-t')^{-\beta}E_{\alpha-\beta, 1-\beta}\qty(-\frac{\mathbb G}{\mathbb V}(t-t')^{\alpha-\beta})\\
    &M(t-t') = -\mathbb G (t-t')^{-1-\beta}E_{\alpha-\beta, -\beta}\qty(-\frac{\mathbb G}{\mathbb V}(t-t')^{\alpha-\beta})
    \label{eq:memory_fractional}
\end{align}
where $E_{a,b}$ is the generalized Mittag-Leffler function, given by
\begin{equation}
    E_{a,b}(z) = \sum_{k=0}^\infty \frac{z^k}{\Gamma(ak+b)}.
\end{equation}
In particular, $G(t)\to0$ for $t\to+\infty$ and $M(t_1)>M(t_2)$ for $t_1<t_2$.

Ultimately, our simulations ground on the following constitutive equation:
\begin{equation}
    \vt \tau(t) = \int_{-\infty}^t
    \!\left[-\mathbb G (t-t')^{-1-\beta}E_{\alpha-\beta, -\beta}\qty(-\frac{\mathbb G}{\mathbb V}(t-t')^{\alpha-\beta})\right]
    \frac{1}{1+a\gamma(t')^b}(\gamma(t')-\gamma(t))
    \dd t'.
    \label{eq:K-BKZ_fractional}
\end{equation}

\subsubsection{Oscillatory shear}
We now consider oscillatory shearing strain input of the form $\gamma(t)=\gamma_0\sin(\omega t)$, with corresponding shear rate $\dot{\gamma}(t)=\omega\gamma_0\cos(\omega t)$. Experiments testing such a flow are of two types, which differ for the magnitude of the amplitude $\gamma_0$ of the sinusoidal oscillations: the \emph{small amplitude oscillatory shear (SAOS)} and the \emph{large amplitude oscillatory shear (LAOS)} experiments. The former is typically conducted at a fixed small amplitude (approximately $\gamma_0\sim10^{-2}$) and measures the amplitude and phase shift of the shear stress as a function of the frequency $\omega$ and allows to characterise the nearly linear behaviour of the fluid. In contrast, in the latter, the oscillations are applied at a fixed frequency with increasing amplitudes and allow quantifying the viscoelastic nonlinear response. Typical values for the strain amplitudes required to go beyond the linear regime are of the order $\gamma_0\sim 10^0$.


In the nonlinear regime, the storage modulus $G'(\omega)$ and the loss modulus $G''(\omega)$ acquire an additional dependency on the input strain amplitude $\gamma_0$ and lose their precise physical meaning of elastic and viscous components. This additional dependency can be easily understood once we consider that these moduli are built from the Fourier transformation of the shear stress response to a sinusoidal strain input, and when large deformations are applied the response is not purely sinusoidal, but rather a superposition of harmonics of higher frequencies.

Therefore, the shear stress output can be represented as a linear superposition of harmonic contributions and be taken of the form \citep{hyun2011review}
\begin{equation}
    \vt\tau(t)=\sum_{p,odd}\sum_{q,odd}^p
    \gamma_0^q\left(a_{pq}\sin(q\omega t)+b_{pq}\cos(q\omega t)\right)
\end{equation}
which is often referred to as a power series \citep{giacomin2011large}. It is important to note that the series contains only the odd harmonics of the shear deformations, and $a_{11}=G'(\omega)$, $b_{11}=G''(\omega)$ are the linear moduli.

An alternative way to represent $\vt \tau(t)$, named Fourier series, reads
\begin{equation}
    \label{eq:fourier_series}
    \vt\tau(t)=\gamma_0\sum_{n,odd}\left(G'_{n}(\omega, \gamma_0)\sin(n\omega t)+G''_{n}(\omega, \gamma_0)\cos(n\omega t)\right)
\end{equation}
and this will be the preferred form in the rest of the manuscript. In this representation, the coefficients $G'_{n}(\omega, \gamma_0)$ and $G''_{n}(\omega, \gamma_0)$ are the Fourier coefficients of the stress response, which depend on both the frequency $\omega$ and the strain amplitude $\gamma_0$. The first harmonic corresponds to the linear moduli: $G'(\omega)=G'_{1}(\omega, \gamma_0)$ and $G''(\omega)=G''_{1}(\omega, \gamma_0)$.

\emph{SAOS limit}.
The linear viscoelastic response of a material described by the fractional K-BKZ model can be obtained by the following procedure.
If we analyse a system described by a single spring-pot and apply a step strain $\gamma(t) = \gamma_0 H(t)$, with $\gamma_0$ sufficiently small, it results in a power-law decay of the stress: 
\begin{equation}
    \tau(t)/\gamma_0 \equiv G(t) = \frac{\mathbb{G}}{\Gamma(1-\beta)}t^{-\beta}
    \label{eq:relax_modulus_singleSP_time}
\end{equation}
which is the response of a class of material known as critical gel \citep{rathinaraj2021incorporating}. The behaviour under SAOS can be computed by Fourier transformation of its constitutive equation, obtaining
\begin{align}
    &G'(\omega) = \mathbb{G}\cos(\pi\beta/2)\omega^\beta\\
    &G''(\omega) = \mathbb{G}\sin(\pi\beta/2)\omega^\beta.
\end{align}
The response of the Fractional Maxwell Model is obtained by the same procedure and results in the following expressions for the linear moduli:
\begin{align}
    &G'(\omega) = \frac{ (\mathbb{G}\omega^\beta)^2{\mathbb V\omega^\alpha}\cos{(\pi\alpha/2)}+(\mathbb{V}\omega^\alpha)^2 \mathbb G\omega^\beta\cos{(\pi\beta/2)}} 
    {(\mathbb{V}\omega^\alpha)^2+(\mathbb{G}\omega^\beta)^2+2\mathbb V \mathbb G \omega^{\alpha+\beta}\cos{[\pi(\alpha-\beta)/2]}}\\
    &G''(\omega) = \frac{ (\mathbb{G}\omega^\beta)^2{\mathbb V\omega^\alpha}\sin{(\pi\alpha/2)}+(\mathbb{V}\omega^\alpha)^2 \mathbb G\omega^\beta\sin{(\pi\beta/2)}} 
    {(\mathbb{V}\omega^\alpha)^2+(\mathbb{G}\omega^\beta)^2+2\mathbb V \mathbb G \omega^{\alpha+\beta}\cos{[\pi(\alpha-\beta)/2]}}
\end{align}

\section{Experimental and numerical setups}\label{sec:experimental}

\subsection{Experimental setup}\label{sec:experimental_setup}
\subsubsection{Materials}

The material selected for the present study was an unreinforced isotactic Polypropylene (iPP) homopolymer, supplied by Borealis (Linz, Austria). The specific grade used was HF955MO, which is a widely used thermoplastic polymer in various industrial applications. Table \ref{tab:material_properties} summarizes the basic mechanical and physical properties of the material, as provided by the resin manufacturer.
Additionally, the material supplier specified an average molecular weight (Mw) of $206$ kDalton.

\begin{table}[htbp]
    \centering
    \begin{tabular}{|c|c|c|}
        \hline
        \textbf{Property} & \textbf{Typical value} & \textbf{Test method} \\
        \hline
        Density & 0.905 g/cm\textsuperscript{3} & ISO 1183 \\
        \hline
        Melt flow rate (230 °C/2.16 Kg) & 20 g/10 min & ISO 1133-1 \\
        \hline
        Tensile modulus (1 mm/min) & 2200 MPa & ISO 527-2 \\
        \hline
        Tensile strain at yield (50 mm/min) & 6 \% & ISO 527-2 \\
        \hline
        Tensile stress at yield (50 mm/min) & 40 MPa & ISO 527-2 \\
        \hline
        Heat deflection temperature B (0.45 MPa) & 115 ºC & ISO 75-2 \\
        \hline
    \end{tabular}
    \caption{Summary of the physical and thermo-mechanical properties of the selected isotactic Polypropylene (iPP).}
    \label{tab:material_properties}
\end{table}

\begin{figure}[b!]
    \centering
    \includegraphics[width=0.8\textwidth]{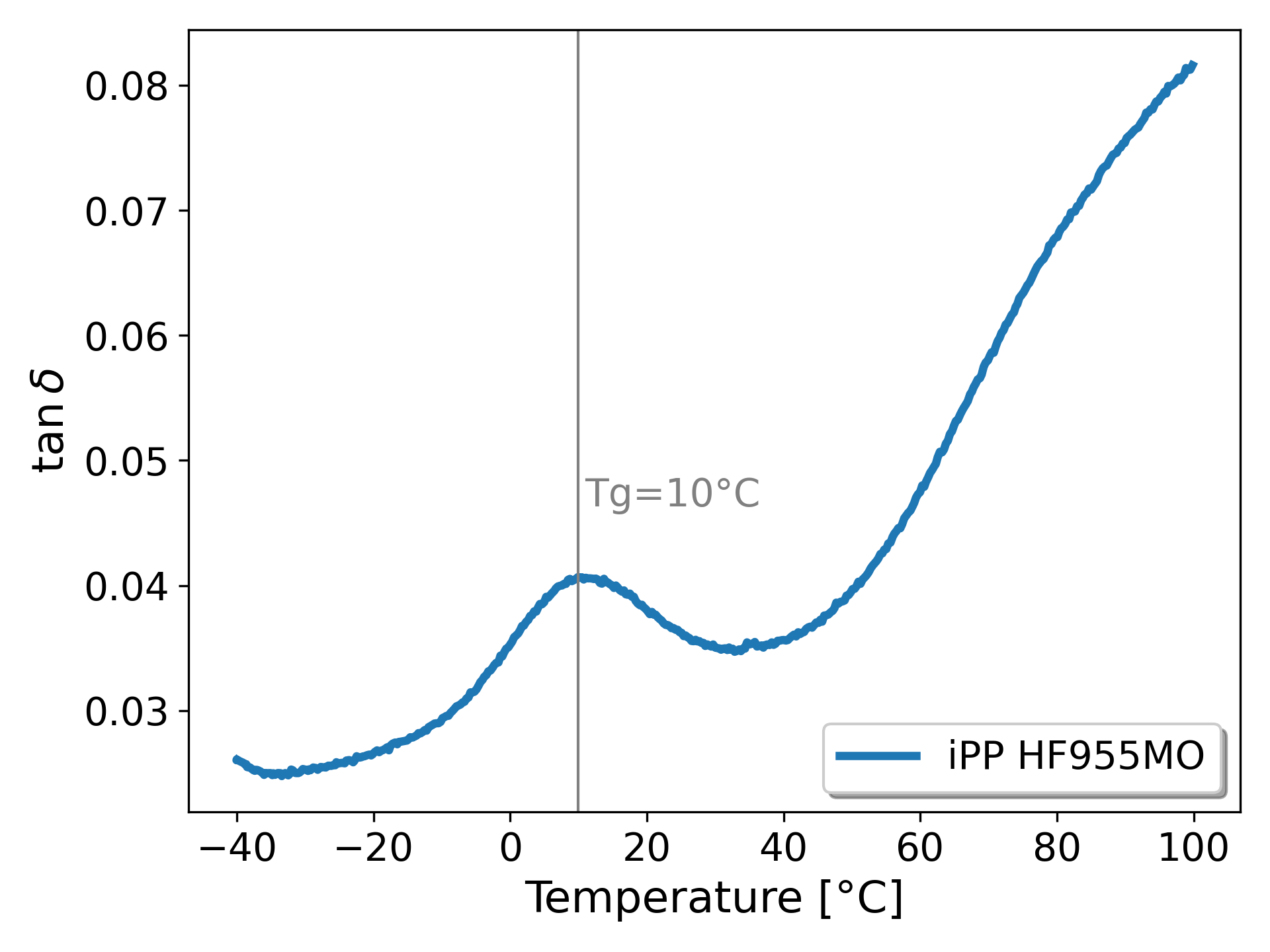}
    \caption{Identification of the $\tan{\delta}$ peak for the iPP material to determine the glass transition temperature (Tg).}
    \label{fig:tandelta_vs_temperature}
\end{figure}

\subsubsection{Experimental methods}

Prior to the melt-state rheological tests, which were the target experiments for further constructing the modelling approach, a series of fundamental measurements were carried out in solid state, i.e. Dynamic Mechanical Thermal Analysis (DMTA) was used to identify the glass transition temperature (Tg) of the iPP, while Differential Scanning Calorimetry (DSC) provided insights into the crystallinity of the injection moulded granules of the material and facilitated the determination of the melting temperature (Tm).

\paragraph{Dynamic Mechanical Thermal Analysis (DMTA) }

A Metravib $+300$ instrument from Acoem (Limonest, France) was used in tensile mode to identify the glass transition temperature (Tg) of the selected Polypropylene. Rectangular coupons, measuring $20$ mm in height and $6$ mm in width, were cut from injection-moulded plaques with a thickness of $2$ mm. The temperature range for the Tg identification was set between $-40$ and $+100$\degree C, with a heating rate of $5$\degree C/min. A frequency of $10$ Hz was applied, in combination with a dynamic displacement amplitude of $5$ micrometers.
Figure \ref{fig:tandelta_vs_temperature} shows the plot of the tangent delta $\tan\delta$, which represents the ratio of the loss modulus to the storage modulus of the material, as a function of temperature. A peak in the $\tan\delta$ values around $10$\degree C corresponds to the Tg of the selected isotactic polypropylene homopolymer. 

\begin{figure}[tbp]
    \centering
    \begin{subfigure}{0.45\textwidth}
        \centering
        \includegraphics[width=\textwidth]{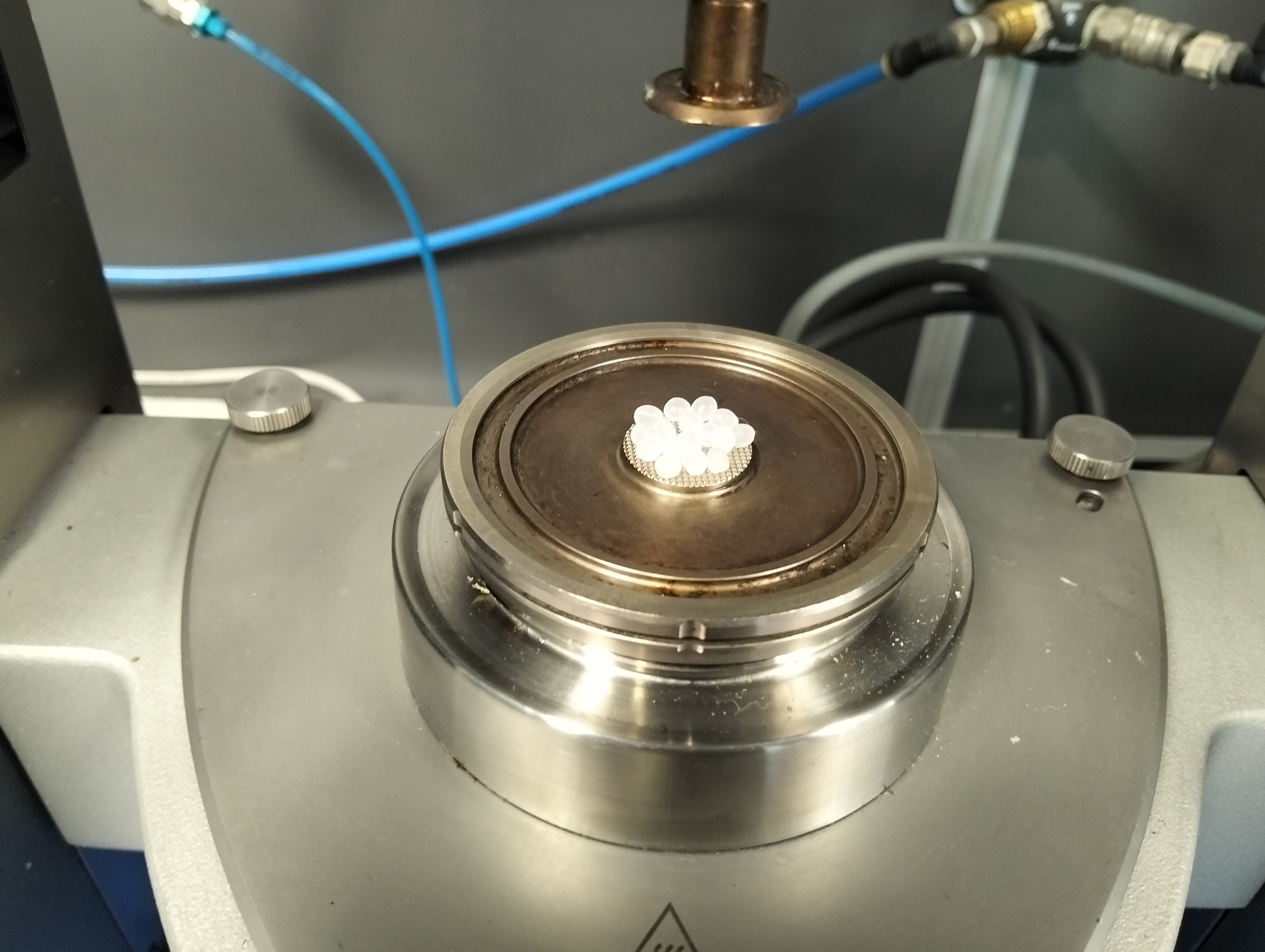}
        \caption{~}
        \label{fig:pellet1}
    \end{subfigure}
    \hfill
    \begin{subfigure}{0.45\textwidth}
        \centering
        \includegraphics[width=\textwidth]{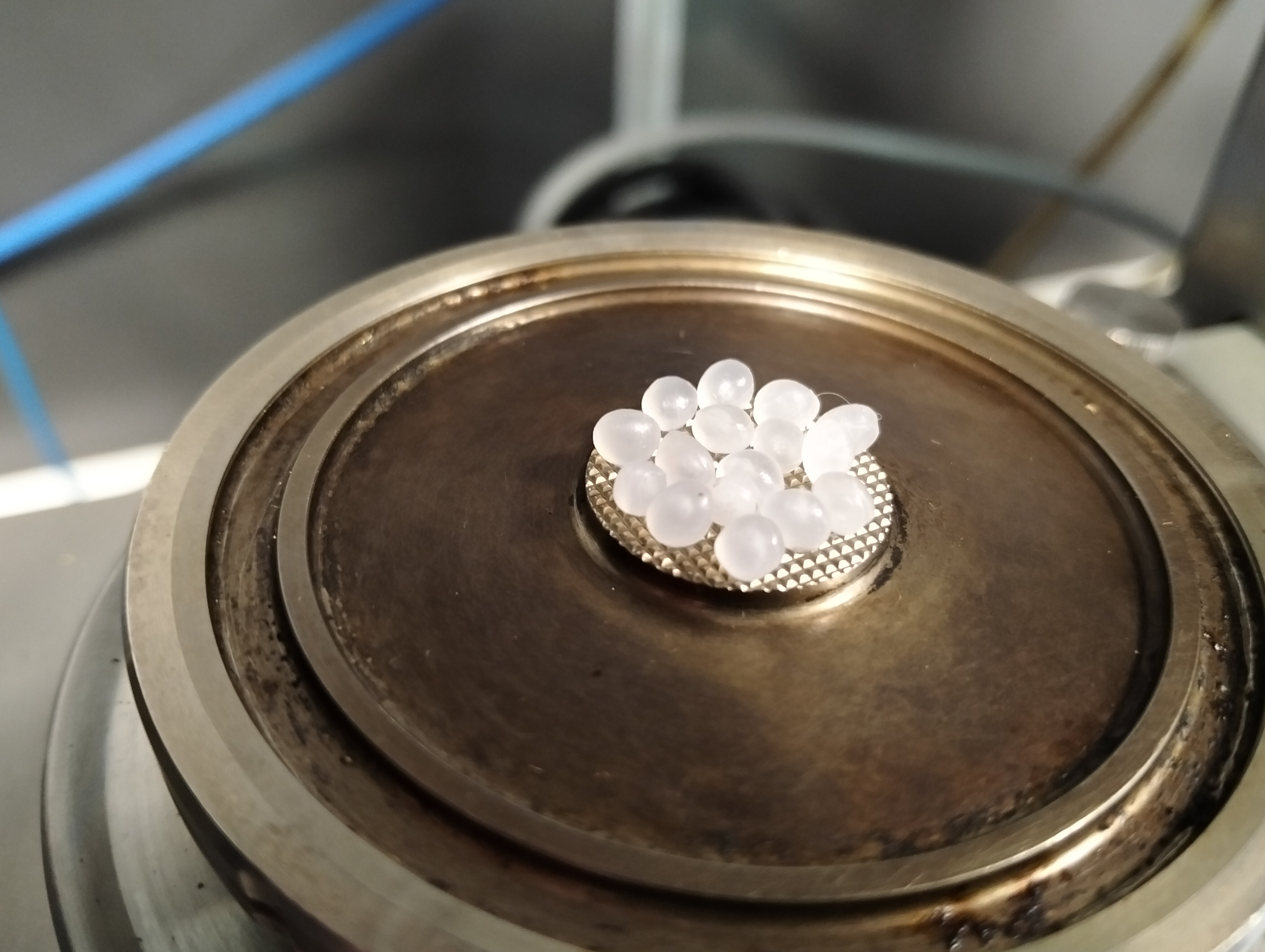}
        \caption{~}
        \label{fig:pellet2}
    \end{subfigure}
    \vskip\baselineskip
    \begin{subfigure}{0.45\textwidth}
        \centering
        \includegraphics[width=\textwidth]{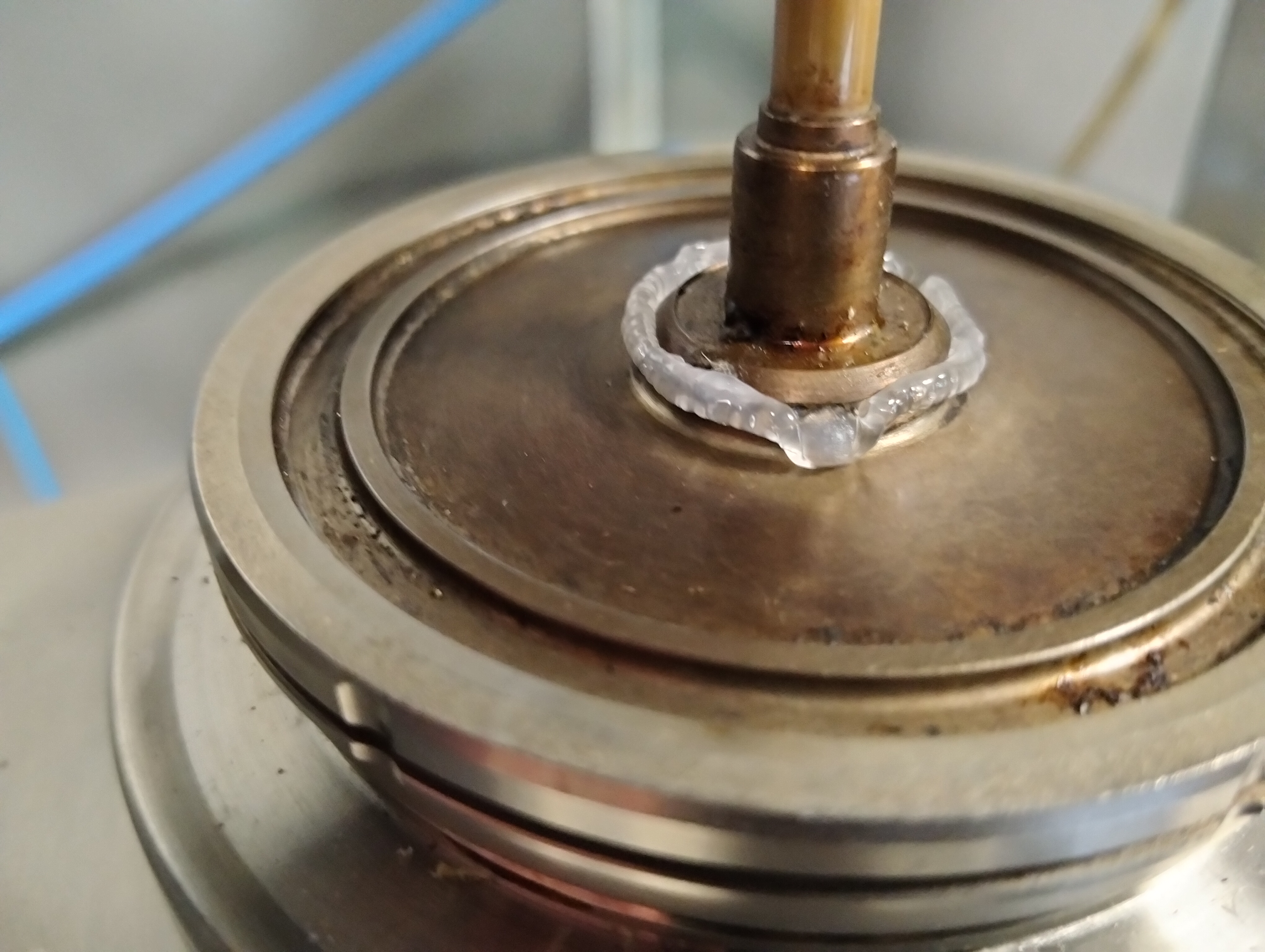}
        \caption{~}
        \label{fig:pellet3}
    \end{subfigure}
    \hfill
    \begin{subfigure}{0.45\textwidth}
        \centering
        \includegraphics[width=\textwidth]{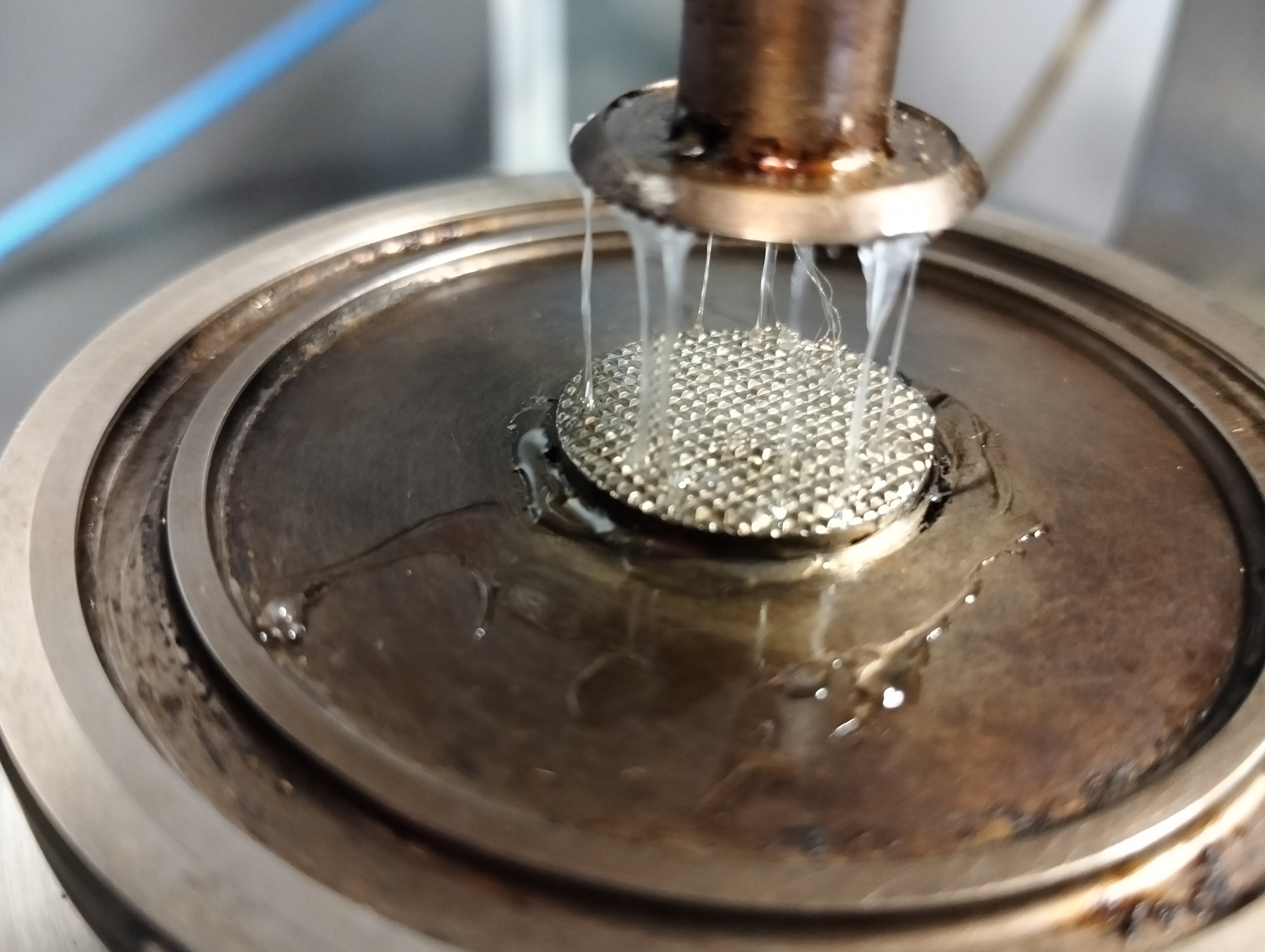}
        \caption{~}
        \label{fig:pellet4}
    \end{subfigure}
    \caption{Material in granular state (a,b) and molten state (c,d) prior to testing in oscillatory mode.}
    \label{fig:pellets}
\end{figure}

\begin{figure}[tbp]
    \centering
    \includegraphics[width=0.7\textwidth]{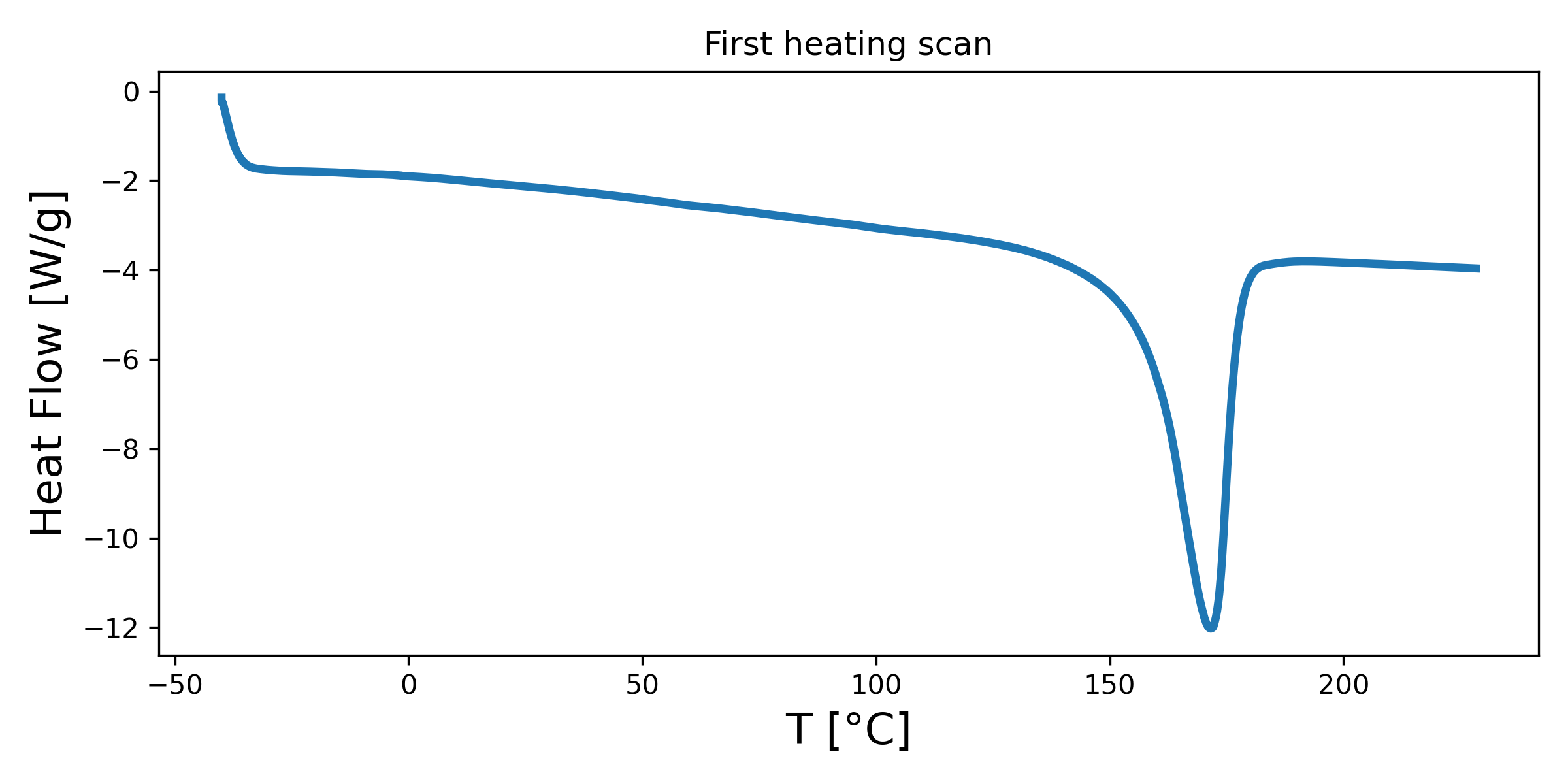}
     \caption{Heat flow versus temperature, identifying $Tm$ at the peak at $171$°C}
    \label{fig:DSC}
\end{figure}

\paragraph{Differential Scanning Calorimetry (DSC)}
A test instrument from TA Instruments (New Castle, USA), model Q 100, was used to determine the melting temperature of the material pellets as well as the crystallinity degree upon cooling. A heating rate of $1\degree C$/min was applied, which resulted in a melting temperature (Tm) of $170\degree C$ and a crystallinity degree of $64\%$. The crystallinity degree was determined using the enthalpy of fusion method, where the heat of fusion of the sample was compared to that of a perfectly crystalline material \citep{lanyi2019determination}. The melting temperature (Tm) was identified as the peak of the endothermic transition in the DSC curve, as shown in Figure \ref{fig:DSC}.

\subsubsection{Melt-state rheology}

The rheological tests performed in the melt state provided the foundational data required for the subsequent modelling of the isotactic polypropylene's flow behaviour. A Haake Mars III torsional rheometer from Thermo Fisher Scientific (Waltham, USA) was used in order to evaluate both the Small Amplitude Oscillatory Shear (SAOS) response and the Large Amplitude Oscillatory Shear (LAOS) regime of the material.

Testing temperature was set to 190°C, above the melting temperature identified by DSC, with material pellets placed on the test specimen lower plate, which had a diameter
of $20$ mm. A gap of $0.5$ mm was employed and all the excess material was trimmed before every measurement. Pictures of the material before and after melting are shown in Figure \ref{fig:pellets}.
The rheometer test campaign for the LAOS measurements consisted of performing strain amplitude sweeps from $20$ to $500\%$ strain, or $0.01$ to $0.25$ radians at a frequency $f_0$ of $1$Hz  ($6.2832$ rad/s), corresponding to a Deborah number for the experimental tests of $De= \lambda f_0 \approx 0.2$, where $\lambda$ is the characteristic relaxation time of the material. The value $\lambda\approx 0.2$ s can be inferred as the inverse of the crossover frequency, as shown in Section \ref{sec:fitting_experimental_data}.
For the SAOS measurements the strain was fixed at $0.1\%$ ($5.10^{-5}$ rad) and the frequencies were varied from $0.1$ to $100$Hz, in order to capture fundamental viscoelastic properties, including shear stiffness ($G'$, $G''$, $G^*$), compliance ($J'$, $J''$, $J^*$), and viscosity ($\eta'$, $\eta''$, $\eta^*$). Additional tests at different frequencies were also conducted to explore the dependency of the response on the chosen $De$ in LAOS conditions. Three different samples were tested for each type of test to ensure a consistent mean response of the material.

\subsection{Numerical implementation}
\label{sec:numerical_implementation}

Simulations of the experiments are implemented using the Smoothed Particle Hydrodynamics (SPH) method, with the structure described in \cite{ellero2010sph,  vazquezquesada2009smoothed} and the integral implementation discussed in \cite{santelli2024smoothed}. The main concepts are summarised here.

SPH  is a mesh-free Lagrangian model discretising the prescribed equations using a set of $N_p$ fluid particles. Each particle represents a discrete fluid element (labelled by $i,j=1,\dots,N_p$). and is associated to its position and momentum.
Mass conservation ensured by defining a particle number density
\begin{equation}
d_i\equiv\sum_j W(|\vb r_{ij}|, r_{cut})
\end{equation}
where (here and thereafter in this section) the sum is intended over all the neighbouring particles of particle $i$. A vector variable with two subscripts is computed as the difference between the values identified by the first and second subscript (e.g. $\vb r_{ij} \equiv \vb r_i - \vb r_j$ is the distance between position $\vb r_i$ and $\vb r_j$).
  $W(|\vb r_{ij}|, r_{cut})$ is an interpolating kernel. The choice of the kernel for this work is a quintic spline kernel function with compact support $r_{cut}$, as in \cite{morris1997modeling}. 
  This choice is an important step in the implementation of SPH, as it affects the accuracy and stability of the numerical method.   Alternatives, such as the cubic spline, the Shepard kernel, or the Wendland kernel, could be used as well; however, they do not possess specific advantages that can be applied in this context \citep{ha2004numerical, yang2014new, chow2018incompressible}. The quintic spline kernel is a popular choice in SPH simulations due to its smoothness and compact support, which has been proved to work correctly for our system \citep{vazquezquesada2019shear}, and we take advantage of its high number of continuous derivatives.
  Finally, the normalised distance value is defined as $\vb e_{ij}=\vb r_{ij}/ r_{ij}$.

Each particle follows the equation of motion (see \cite{espanol2003smoothed})
\begin{align}
    \vb{\dot{r}}_i&= \vb v_i\\
      m\dot{\vb v_i} &= -\sum_j  \qty[\frac{\vc \pi_i}{d_i^2}+ \frac{\vv{\pi}_j }{d_j^2}]W'_{ij}\vb e_{ij}+2\eta_s\sum_j\frac{1}{d_id_j}\frac{W'_{ij}}{r_{ij}} \vb v_{ij}
    \label{eq:SPH_vel}
\end{align}
Here, $\eta_s$ is the viscosity of the Newtonian solvent 
and $W'_{ij}\equiv\pdv{W(\vb r,r_{cut})}{\vb r} \eval_{\vb r = \vb r_{ij}}$ is the magnitude of the gradient of the smoothing kernel.
$\vc \pi_i$ is the total stress tensor, which can be separated into  the extra-stress $\vt \tau_i$ and the isotropic particle pressure tensor $P_i\vt{1}$, where $P_i = p_0[(\rho_i/\rho_0)^\gamma-1]$ and $\rho_0$ and $p_0$ are reference fluid density and pressure respectively while $\gamma=7$ and $c_s=\sqrt{\gamma \frac{p_0}{\rho_0}}$ is the speed of sound of the solvent \cite{monaghan1994simulating}. Following the weak compression approximation, the value of $c_s$ can be chosen to keep density variations small enough.
The extra stress $\vt \tau_i$ is computed by discretising equation \eqref{eq:K-BKZ_fractional}, as described in the next section. Regarding the boundary conditions, we implemented Newtonian boundary conditions for stress as discussed in \cite{vazquezquesada2012spha}. 

\subsubsection{Integral-form implementation of the stress}
The position of each particle is readily accessible in SPH, therefore -to account for periodic boundary conditions- we compute the displacement of a particle compared to the position it had in the past at time $t'=t-n\Delta t$, with  $\Delta t$ time step, as  $\Delta \boldsymbol{r}_{i,n} = \boldsymbol{r}_i(t) - \boldsymbol{r}'_i(t-n \Delta t)$. It updates as 
 \begin{eqnarray}
  \Delta \boldsymbol{r}_{i,{n+1}} =
  \Delta \boldsymbol{r}_{i,n} + 
  \boldsymbol{r}_i(t+\Delta t) - 
  \boldsymbol{r}_i(t).
  \label{eq:update_displacement}
 \end{eqnarray}

 The displacement tensor 
  \begin{eqnarray}
    \vb \Delta_i \equiv \qty(\grad \vb r ')_i = -\frac{1}{d_i}\sum_jW'_{ij}\vb e_{ij}\vb r'_{ij} 
\end{eqnarray}
is then inverted to compute $\vb E_i$ and $\vt B_i = \vb E_i \vb E^{-T}_i$, so that $\vt \gamma_{[0],i} = \vt 1 - \vt B_i$,
leading to a discretised stress 
\begin{equation}
    \vt \tau_i = \int_{t_0-T_{save}}^{t_0} M(t-t')h(\gamma)\vt \gamma_{[0],i}(t,t') \dd t' .
    \label{stress_calculation}
\end{equation}
which is then efficiently computed over $N_s$ points according to the method described in the supplementary material of \cite{santelli2024smoothed}.

\subsubsection{Numerical parameters}\label{sec:numerical_setup}

The setup for the numerical experiments consists of the following. The fluid is confined between two parallel horizontal plates, at a distance $L=7.5$. Periodic boundary conditions are considered for the vertical walls, while no-slip conditions are applied on the horizontal walls. The symmetry of the problems analysed here allow for the simulation to be carried out in a two-dimensional simulation. Thus, the flow domain is set as a square $[0, \, L] \times [0, \, L]$. 
The fluid is initially at rest.
The bottom plate is fixed, while the top plate oscillates at a frequency $\omega_0$ and amplitude $\gamma_0=\Delta/L$, i.e.  the ratio between the maximum displacement of the upper wall $\Delta$ and $L$. Therefore, for transient shear rate $\gamma(t)$ and strain rate $\gammadot(t)$, we have:
\begin{equation}
    \gamma(t) = \gamma_0 \sin(\omega_0 t)
\end{equation}
\begin{equation}
    \gammadot(t) = \gammadot_0 \cos(\omega_0 t),
\end{equation}
with $\gammadot_0 = \omega_0 \gamma_0 = V_{max}/L$. 

The
number of fluid particles in the system is $N_p = N \times N$: since the domain
is a square, we use the same number of particles in the two dimensions.
Spatial resolution has been previously verified by running simulations at various
$N$, fixing the cut-off radius to $r_c=4L/N$ (see also \cite{santelli2024smoothed}), and we chose $N=60$ to keep the error below $5\%$, resulting in a cut-off radius of $r_c=0.5$. 

\begin{figure}
    \centering
    \includegraphics[width=0.7\textwidth]{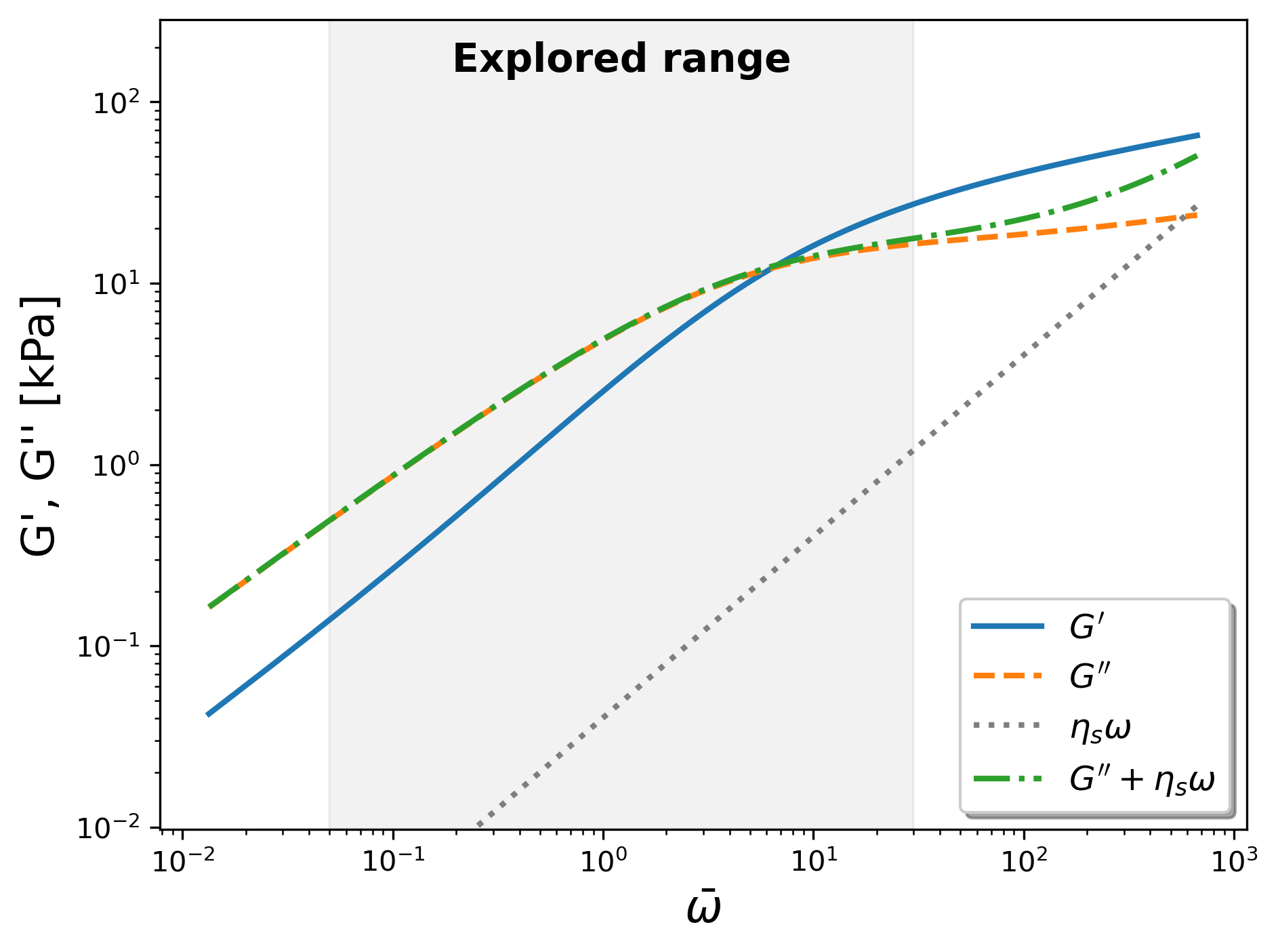}
     \caption{Storage and loss moduli as a function of $\bar\omega=\omega\lambda$. The explored range of frequencies is indicated by a grey shaded area. $G'(\bar\omega)$ ($\textcolor{blue}{-}$) and $G''(\bar\omega)$ ($\textcolor{orange}{- -}$) from the fractional K-BKZ model with parameters $\alpha=0.85$, $\beta=0.2$, $\lambda_c=0.2$ s, $G_c=432$ kPa. For comparison, the artificial Newtonian contribution ($\textcolor{black}{:}$) and the combined behaviour of the fractional model and the artificial Newtonian contribution ($\textcolor[rgb]{0,0.3,0}{-.}$) are shown.}
    \label{fig:artificial_newtonian}
\end{figure}

In order to ensure numerical stability, the scheme benefits from an additional Newtonian viscosity $\eta_s$, which is set to a small  value, much smaller than the viscosity of the fluid. To show that this does not disrupt the dynamics, we can consider the following cases. For small $\omega$, the behaviour is dominated by the asymptotic value of the fractional model $G''(\omega)\sim \omega^\alpha$. For large frequencies, the artificial Newtonian contribution $\eta_s \omega$ is larger than the asymptotic value of the fractional model $G''(\omega)\sim \omega^\beta$, since $\beta<\alpha<1$. Nevertheless, if the value of $\eta_s$ is small enough, the crossover between the two regimes happens at a frequency $\omega_c$ much larger than the frequencies of interest. As shown in Figure \ref{fig:artificial_newtonian}, we have ensured that the artificial Newtonian contribution is always at least one order of magnitude smaller than the loss modulus of the fractional model at the highest frequency of interest.

The characteristic time scale is set upon the the elastic time $\lambda = 1$ (or its closer equivalent when the fractional parameters are outside of the Maxwell regime, see \cite{jaishankar2013power,jaishankar2014fractional}), which ensures that the elastic forces are dominant over the viscous ones and satisfies the series of disequalities $t_{\eta_s}=\frac{L^2\rho}{\eta_s}<{\lambda},\frac{2\pi}{\omega_0}$, where $t_{\eta_s}$ is the viscous time, while $2\pi/\omega_0$ is the period of the oscillations, fixed at a angular frequency $\omega_0$. For most of the simulations,  $\omega_0=\lambda^{-1}$. 
The time step is set to $\Delta t \approx 10^{-5}\lambda$. Regarding the parameters of the  numerical implementation of the integral model, we have $T_{save}=10\lambda$ and $N_s=100$. 

Dimensionless numbers are defined as follows: Reynolds number $Re=\frac{\rho V_{max} L}{\eta_{tot}}$, Weissenberg number $Wi=\frac{\lambda V_{max}}{L}$, Deborah number $De=\frac{\lambda}{2\pi/\omega_0}$. The simulations are performed, when not differently specified, for $De \approx 0.2$.

\begin{figure}[tbp]
    \centering
    \includegraphics[width=0.7\textwidth]{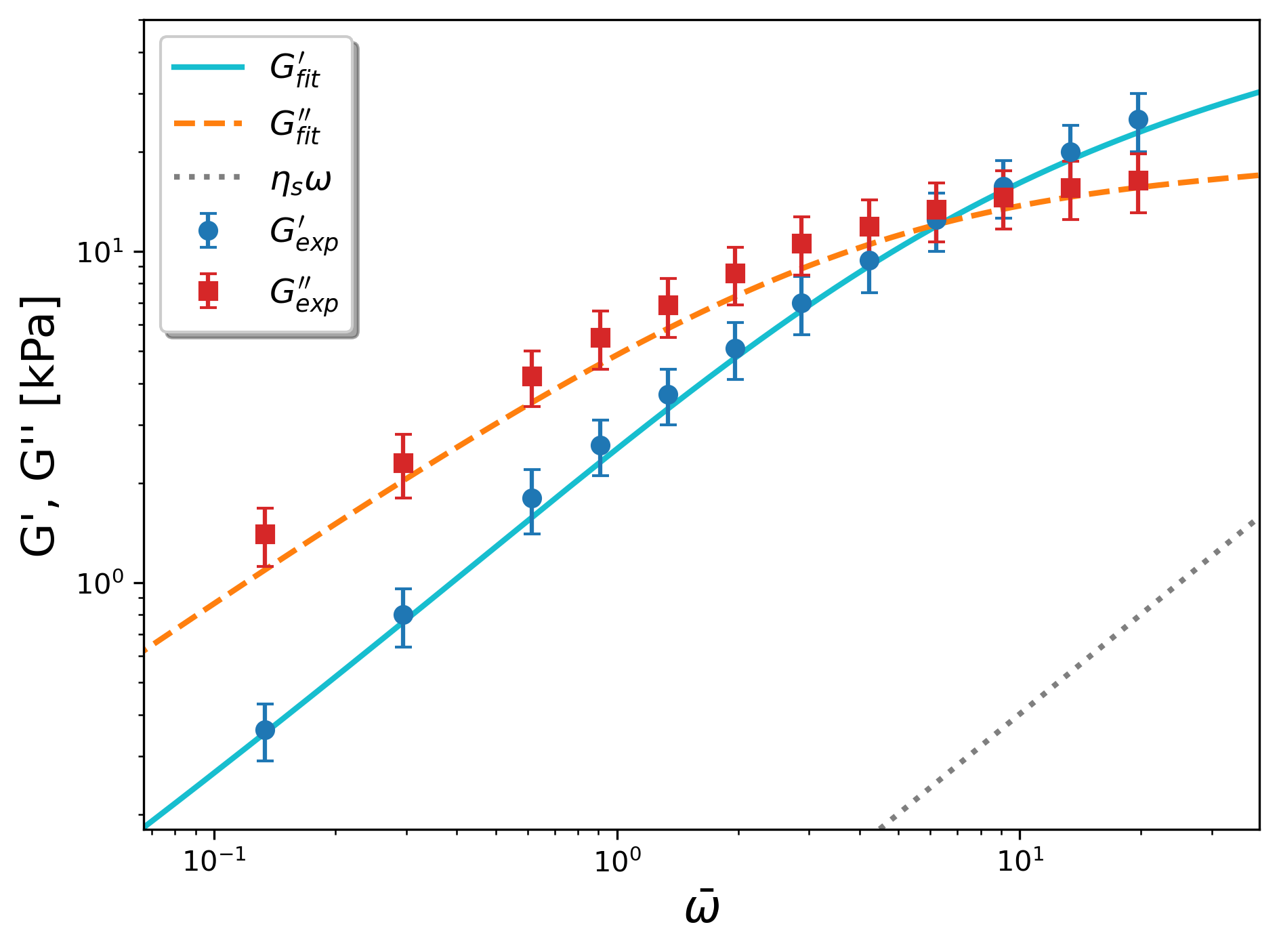}
     \caption{Storage and loss  moduli from SAOS as a function of $\bar\omega=\omega\lambda$: experimental data (\textcolor{blue}{$\bullet$} and \textcolor{red}{$\scalebox{0.6}{$\blacksquare$}$} symbols), fractional K-BKZ model (respectively \textcolor{cyan}{\rule[0.5ex]{0.8em}{0.8pt}} and \textcolor{orange}{\rule[0.5ex]{0.4em}{0.8pt}~\rule[0.5ex]{0.4em}{0.8pt}}) with parameters $\alpha=0.85$, $\beta=0.2$, $\lambda_c=0.2$ s, $G_c=432$ kPa. Artificial Newtonian contribution (\textcolor{black}{:}) shown for comparison.}
    \label{fig:comparison_SAOS}
\end{figure}

\section{Results}\label{sec:results}

In this section we analyse the experimental data obtained from the rheological tests, and use these results to calibrate the parameters of the fractional K-BKZ model. We then compare the experimental data with the numerical results obtained from the SPH simulations. 
\subsection{Fitting of experimental data}
\label{sec:fitting_experimental_data}
First, we focus on the choice of the fractional parameters required to capture the experimental data. The initial fractional parameters requiring calibration are better identified under linear conditions, i.e. small amplitude oscillatory shear (SAOS). Therefore we have performed a series of SAOS tests, where the strain amplitude is kept small enough to ensure that the material behaves linearly and $Re$ is always below $0.01$. The results of these tests are shown in Figure \ref{fig:comparison_SAOS}, where we compare the experimental data obtained from runs at $\gamma_0<0.1$ with the numerical predictions of the fractional K-BKZ model under the assumption of small displacement. From the same data we can also infer the characteristic relaxation time of the material, which is $\lambda\approx0.2$ s, as the inverse of the crossover frequency between $G'$ and $G''$.
 We have observed a good agreement for parameters $\alpha=0.85$ and $\beta=0.2$, resulting in a sample with a fractional behaviour close but distinct from a classical Maxwell model, which would correspond to $\alpha=1$ and $\beta=0$. The characteristic modulus and time are set to $G_c=432$ kPa and $\lambda_c=0.2$ s respectively.
 
\begin{figure}[b]
    \centering
    \begin{subfigure}{0.48\textwidth}
        \centering
        \includegraphics[width=\textwidth]{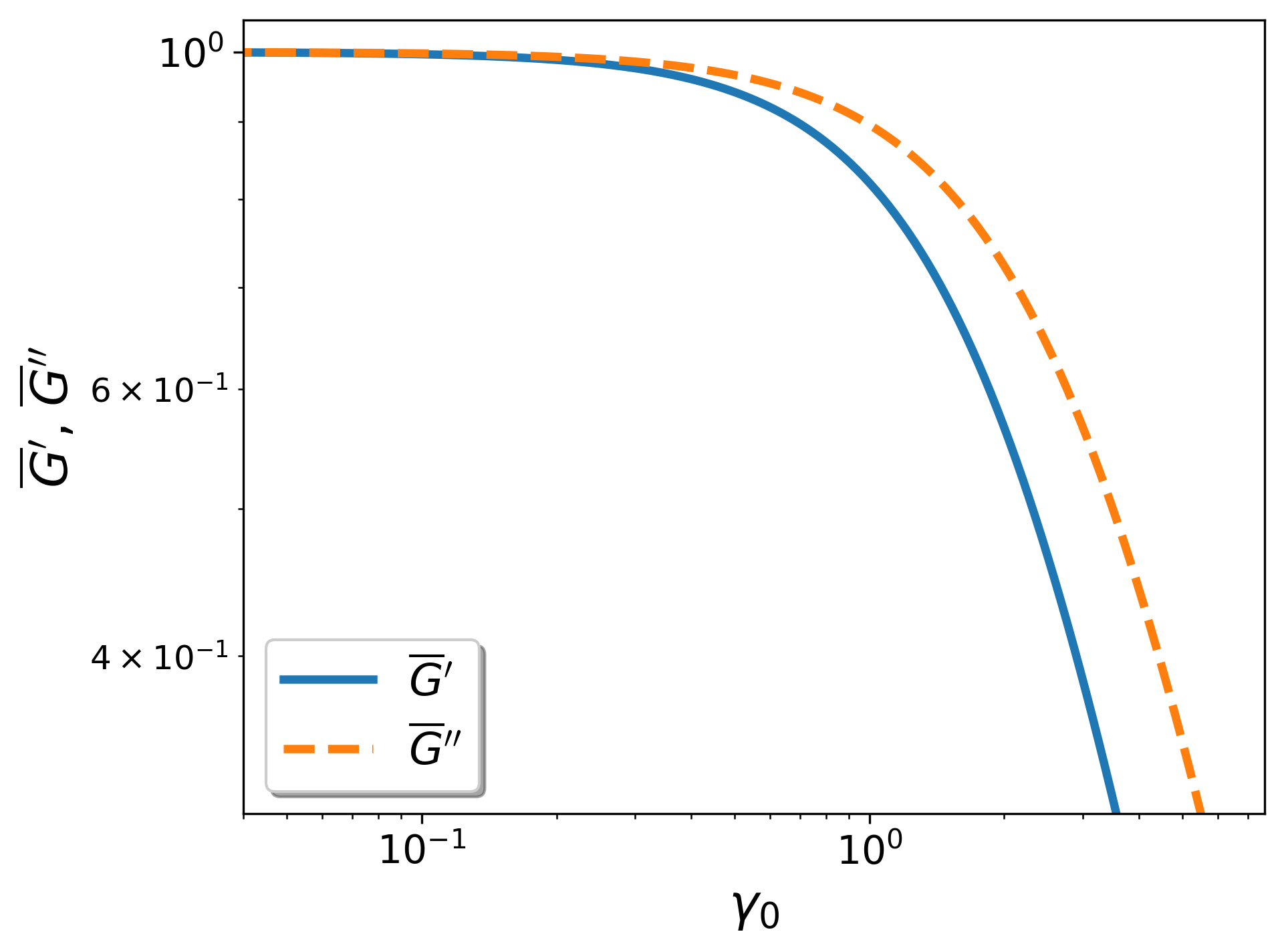}
        \caption{Normalised $G'$ and $G''$ (\textcolor{blue}{\rule[0.5ex]{0.8em}{0.8pt}} and \textcolor{orange}{\rule[0.5ex]{0.4em}{0.8pt}~\rule[0.5ex]{0.4em}{0.8pt}}) as a function of strain amplitude $\gamma_0$. The curves are extrapolated from experimental data}
        \label{fig:G12_thinning}
    \end{subfigure}
    \hfill
    \begin{subfigure}{0.48\textwidth}
        \centering
        \includegraphics[width=\textwidth]{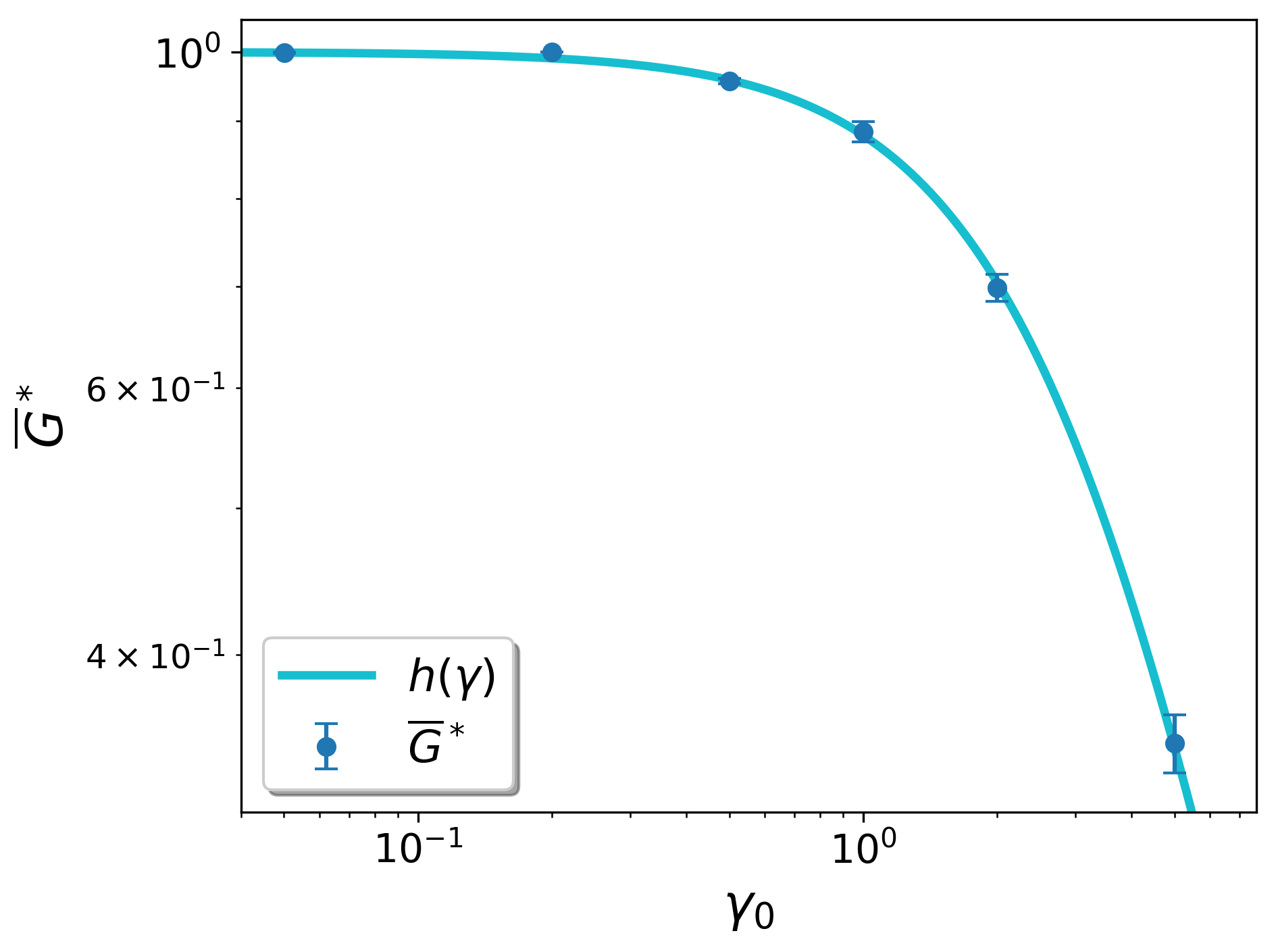}
        \caption{Normalised $G^*$ as a function of strain amplitude $\gamma_0$. Symbols for experimental data, solid line for fitted $h(\gamma)$. } 
        \label{fig:Gstar_thinning}
    \end{subfigure}
    \caption{Analysis of experimental data at fixed $De\approx0.2$, showing strain thinning.}
    \label{fig:experimental_G}
\end{figure}

\begin{figure}[tbp]
    \centering
    \begin{subfigure}{0.49\textwidth}
        \centering
        \includegraphics[width=\textwidth]{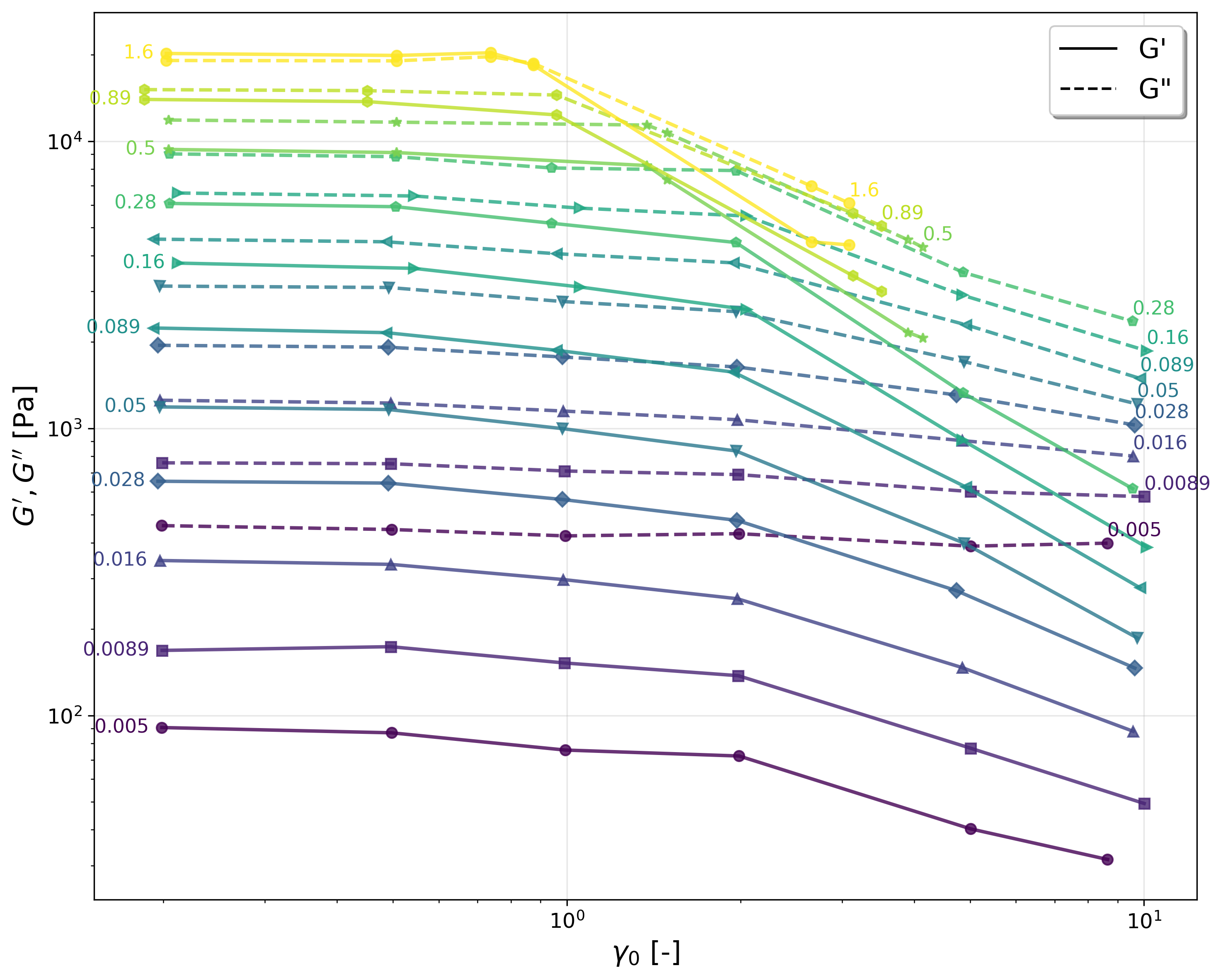}
        \caption{Storage and loss moduli vs. strain amplitude $\gamma_0$.}
        \label{fig:combined_G_vs_gamma}
    \end{subfigure}
    \hfill
    \begin{subfigure}{0.49\textwidth}
        \centering
        \includegraphics[width=\textwidth]{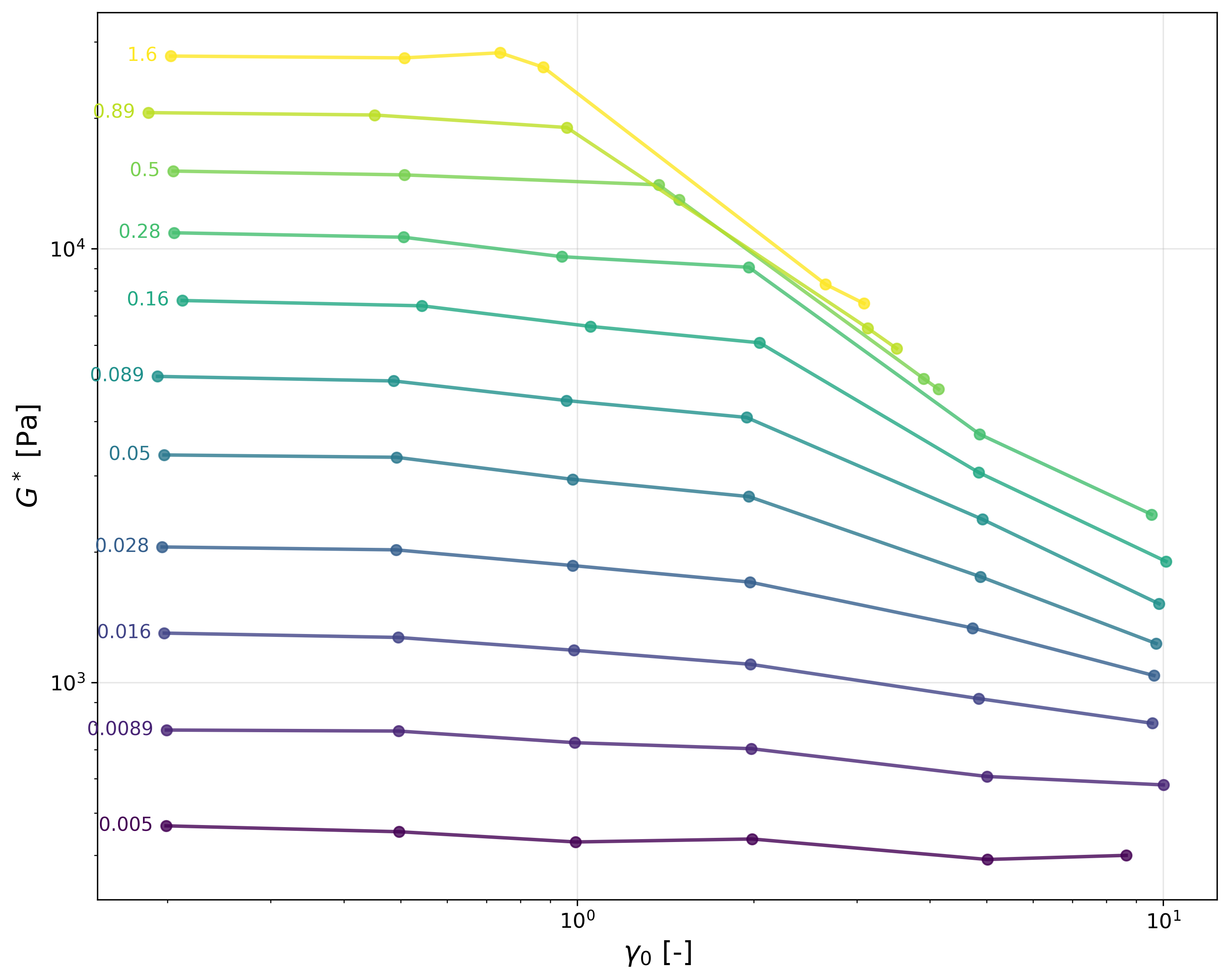}
        \caption{Complex modulus $G^*$ vs. strain amplitude $\gamma_0$.}
        \label{fig:complex_modulus_G_star}
    \end{subfigure}
    \caption{Left: $G'$ (-) and $G''$ (- -) as a function of strain amplitude $\gamma_0$. Right: complex modulus $G^*$ as a function of $\gamma_0$. The numbers next to the symbols indicate the value of $De$ for each series of experimental data.}
    \label{fig:strain_sweep_De}
\end{figure}

 Afterwards, we focus on the implementation of the non-linear behaviour, where the required function to calibrate is the damping function $h(\gamma)$, defined in equation \eqref{eq:damping_function}. We gathered experimental data at fixed frequency $f_0=1$~s$^{-1}$ for two order of magnitudes of values of initial strain $\gamma_0$. In Figure \ref{fig:experimental_G}, the values of $G'$, $G''$ and the relaxation modulus $G^*$ are normalised over their respective value at $\gamma_0\to0$
 (respectively, $G'(0)=3.62\text{ kPa}, G''(0)=6.88\text{ kPa}, G^*(0)=7.77\text{ kPa}$ at $De\approx 0.2$)
 . Figure \ref{fig:G12_thinning} clearly shows that the polypropylene we analysed behaves as a strain thinning fluid for $De\approx0.2$ \citep{hyun2011review}. Therefore, following the approach of \cite{jaishankar2014fractional}, we obtained the fitting parameters for the damping function (see Figure \ref{fig:Gstar_thinning}), which results in $h(\gamma)=(1+0.2\gamma^{1.9})^{-1}$ (i.e. $\gamma^*\approx2.3$).
 Henceforth, the parameters of the numerical scheme are set to the values obtained from the fitting of the experimental data.
 A strain sweep for multiple values of $De$ is shown in Figure \ref{fig:strain_sweep_De}, where $G'$, $G''$ and $G^*$ are plotted as a function of $\gamma_0$. The results show that the strain thinning behaviour is more pronounced for larger values of $De$, as expected.

\subsection{LAOS analysis}
\label{sec:LAOS_analysis}

\begin{figure}[bpt]
    \centering
    \begin{subfigure}{0.48\textwidth}
        \centering
        \includegraphics[width=\textwidth]{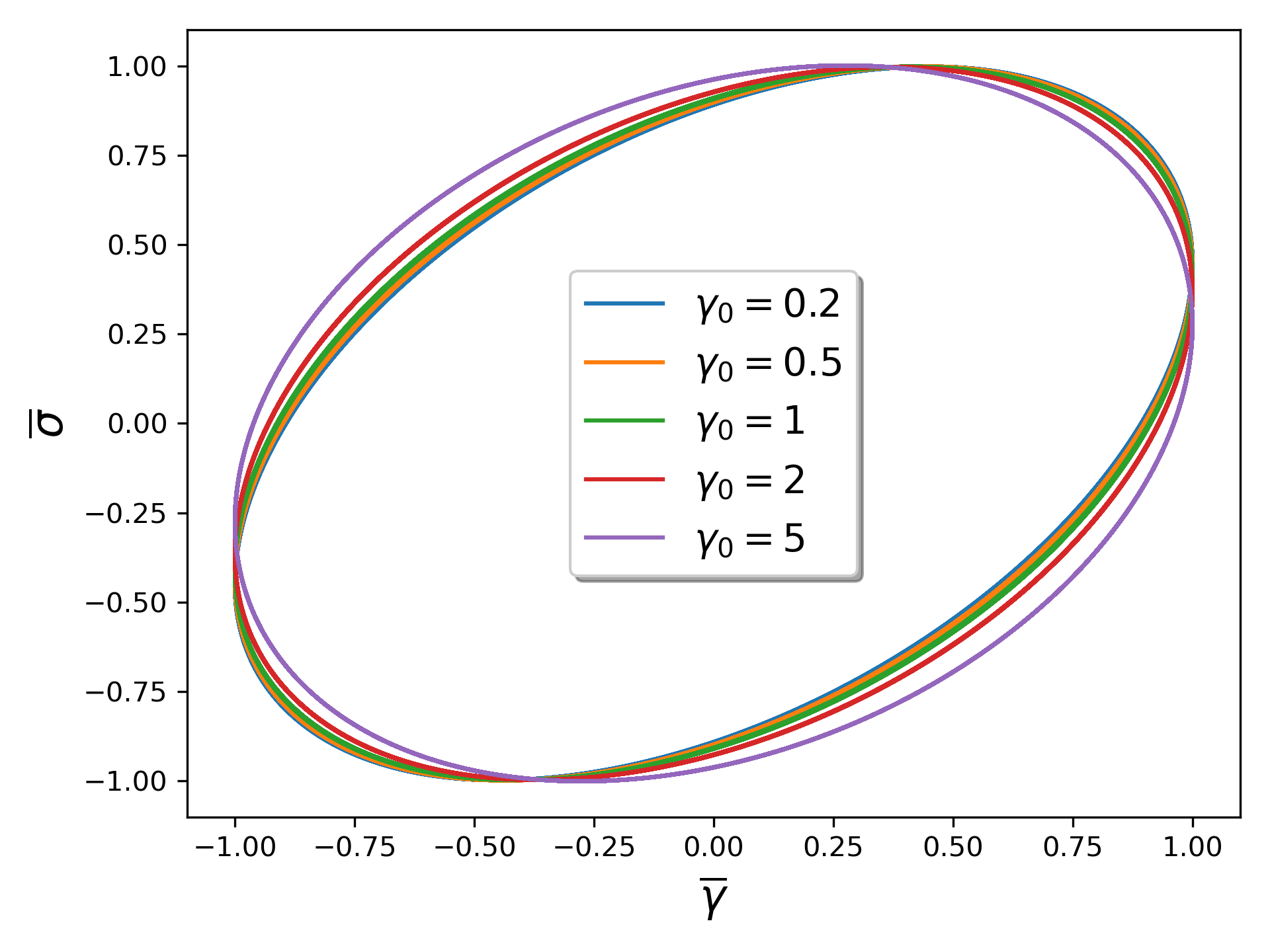}
        \caption{Lissajous curves for multiple $\gamma_0$.}
        \label{fig:june1_lissajous_all}
    \end{subfigure}
    \hfill
    \begin{subfigure}{0.48\textwidth}
        \centering
        \includegraphics[width=\textwidth]{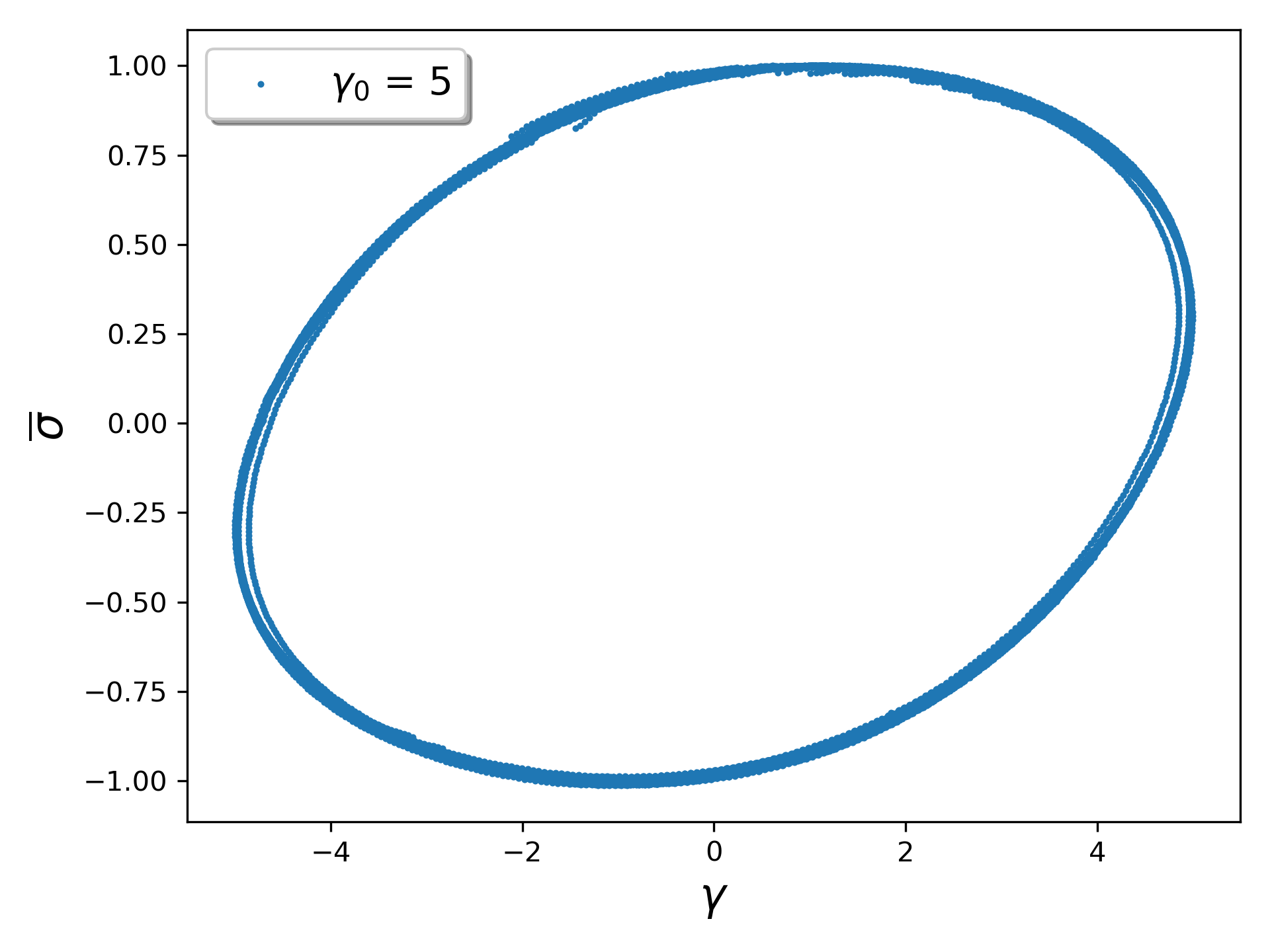}
        \caption{Lissajous curve for large $\gamma_0$.}
        \label{fig:june1_lissajous_4}
    \end{subfigure}
    \caption{Elastic Lissajous curves from SPH simulations at $De\approx 0.2$ for normalised stress $\overline{\sigma}=\tau/\tau_{max}$: (a)  average curves for $0.2<\gamma_0<5$, plotted versus normalised shear $\overline{\gamma}=\gamma/\gamma_{0}$, (b) details for $\gamma_0=5$. The non-elliptical shape is an indicator of nonlinear viscoelastic behaviour.}
    \label{fig:june1_lissajous}
\end{figure}

The appearance of non-linearity can be observed from multiple points of view. Plotting the elastic and viscous Lissajous curves produces patterns that deviate from ellipses, but rather closed loops, as shown in Figure \ref{fig:june1_lissajous}, where the normalised stress $\tau(\gamma)/\tau_{max}$ is plotted versus the shear $\gamma(t)$ or the normalised shear $\overline{\gamma(t)}=\gamma(t)/\gamma_{0}$. However, this approach struggles in quantifying the non-linearity, and is especially hard to apply when the non-linear behaviour is just emerging. Indeed, it can be noticed from Figure \ref{fig:june1_lissajous_all} that the Lissajous curves for small to medium $\gamma_0$ are almost indistinguishable from the linear case, with the non linearity clearly appearing only for $\gamma_0=5$.

A quantitative analysis of the results requires the application of a Fourier transform to the stress signal, in order to obtain the generalised storage and loss moduli. A very deep and clarifying introduction to this approach is provided by \cite{wilhelm2002fourier}, where the most common techniques and pitfalls are discussed.
Detailed and systematic approaches to the analysis of LAOS data using the Fourier analysis can be found in \cite{ewoldt2008new,ewoldt2008ontology} as well as in \cite{hyun2003nonlinear,ng2011large,dimitriou2013describing}. We emphasise that the analysis following the Fourier transform is not the only possible approach to analyse LAOS data: \cite{rogers2011sequence,rogers2012sequence_a,rogers2012sequence_b}, for example, proposed a sequence of physical processes (SPP) method for interpreting the physics behind a LAOS signal, which might provide additional insights compared to the Fourier approach \citep{erturk2022comparison}. However, we will focus on the Fourier approach in the following, as it provides a straightforward way to quantify the non-linearities appearing in the stress signal and a solid framework for further analysis.

In particular, \cite{ewoldt2008new} discuss the various definitions of the non-linear storage and loss moduli that can be obtained from the Fourier transform, and the advantages of describing such systems using $e_i$ and $v_i$, the elastic and viscous Chebyshev coefficients, which can be obtained from the Fourier coefficients. In particular, if we decompose the stress signal as equation \eqref{eq:fourier_series}, repeated here for clarity,
\begin{equation}
    \tau(t;\omega,\gamma_0) = \gamma_0\sum_{n=\text{odd}} \left\{ G'_n(\omega,\gamma_0) \sin(n\omega t) + G''_n(\omega,\gamma_0) \cos(n\omega t) \right\}
\end{equation}
we can compute the Chebyshev coefficients as
\begin{equation}
    e_n = G'_n (-1)^{\frac{n-1}{2}} \quad , \quad v_n = \frac{G''_n}{\omega}
\end{equation}
The advantage of using these coefficients is that they  can be directly related to the intra-cycle non-linearity of the material. In particular, the third order coefficients $e_3$ and $v_3$ are good measures of the intra-cycle strain-stiffening/softening and shear-thickening/thinning behaviour of the material, respectively \citep{ewoldt2008new}. We also note that the Fourier transform of the stress can also be written in terms of intensity $|G^*|$ and phase $\delta$ of each harmonic, as
\begin{equation}
    \tau(t;\omega,\gamma_0) = \gamma_0\sum_{n=\text{odd}} |G^*_n(\omega,\gamma_0)| \sin(n\omega t + \delta_n)
\end{equation}
with $G'_n = Re(G^*_n)$ and $G''_n = Im(G^*_n)$.

Following the guidelines outlined in \cite{hyun2011review}, we have chosen to avoid a simple Fast Fourier Transform (FFT) butterfly method, as it may induce shifts in the transformed signal. Therefore, we applied a Direct Fourier Transform (DFT) to the stress signal for the experimental data we have available, which is a more robust method for obtaining the Fourier coefficients. Nevertheless, the amount of data points we have available for the simulated data is often above the recommended limit of DFT, so -after verifying that the results are compatible between the two methods- we have chosen to apply a pseudo-periodic FFT, where, assuming $f(t_1-t_0)$ is the zero-mean signal between two zeros, we construct a longer signal iteratively as $f(t_{n+1}-t_n)=-f(t_{n}-t_{n-1})$, where $t_n$ is the time of the $n$-th zero. The resulting signal is a periodic signal of the chosen length, which is then transformed using the FFT method. 

\begin{figure}[tbp]
    \centering
    \begin{subfigure}{0.495\textwidth}
        \centering
        \includegraphics[width=\textwidth]{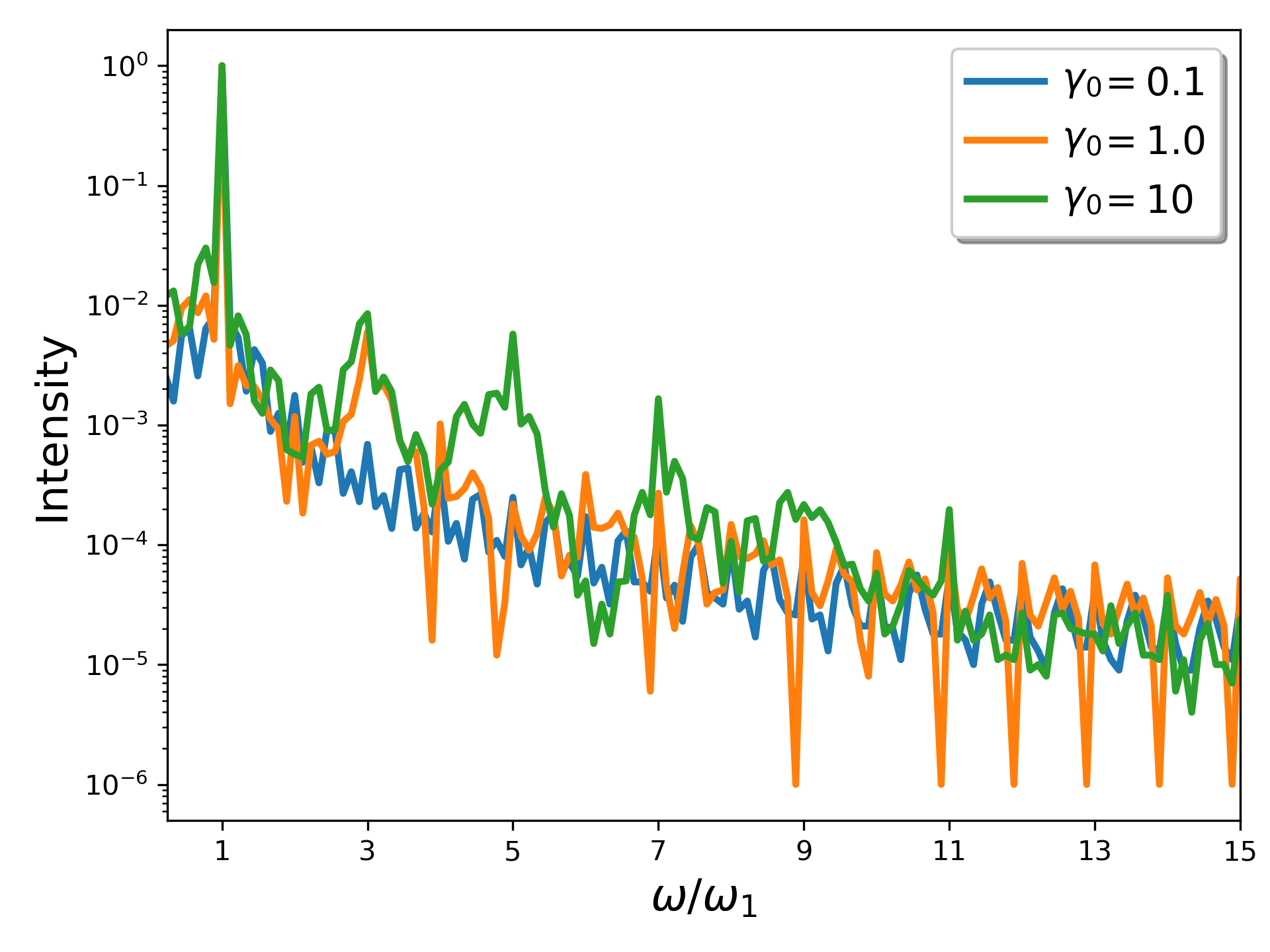}
        \caption{Intensity normalised over the first harmonic.}
        \label{fig:SPH_fourier_peaks}
    \end{subfigure}
    \hfill    
    \begin{subfigure}{0.495\textwidth}
        \centering
        \includegraphics[width=\textwidth]{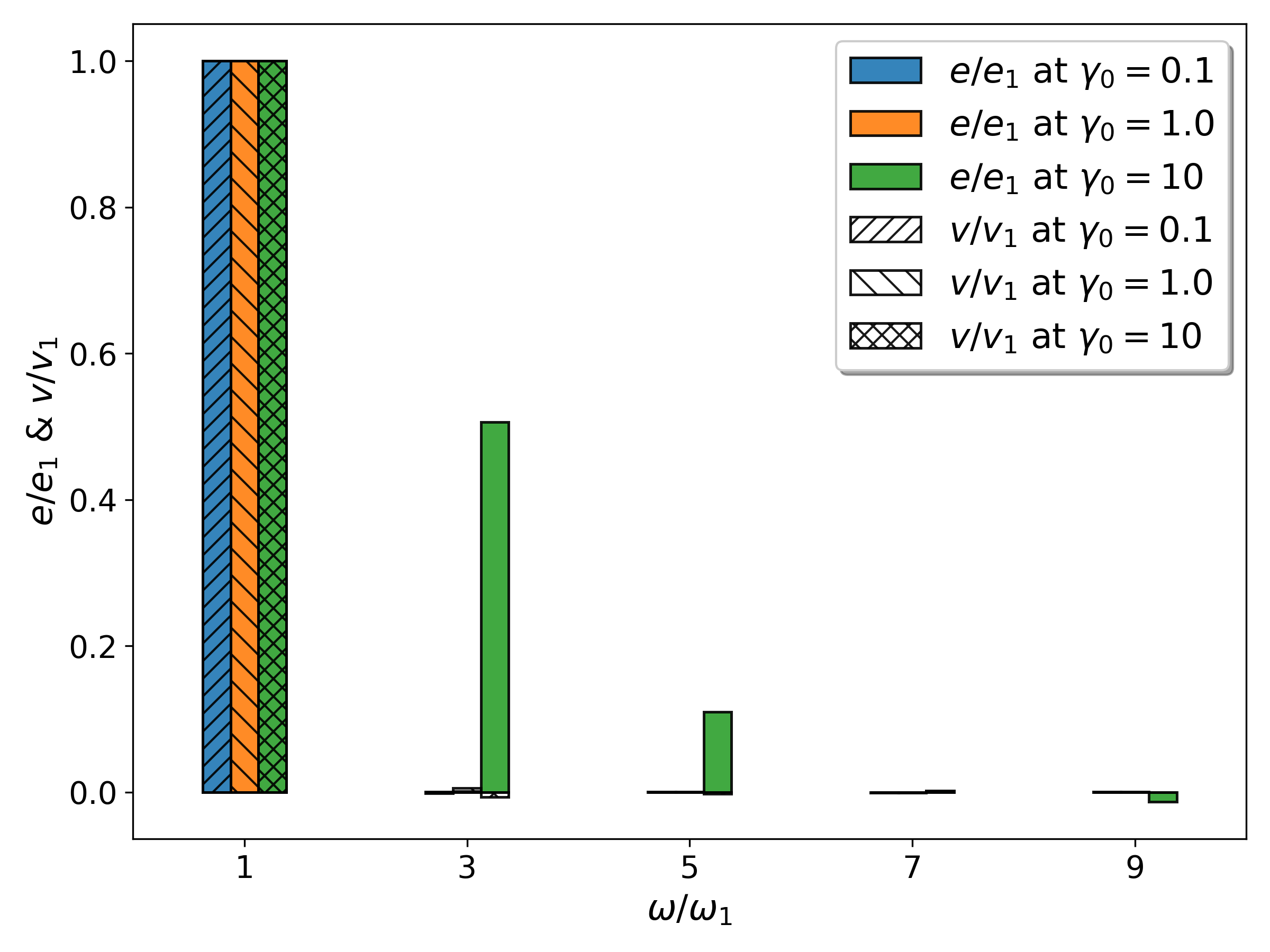}
        \caption{Chebyshev coefficients normalised over the first harmonic.}
        \label{fig:chebyshev_coefficients}
    \end{subfigure}
    \caption{Fourier transform of the stress signal for various values of the strain amplitude $\gamma_0$. Higher harmonics appear only for large displacements, and their relative intensity (compared to the first harmonic) and amount increase with the strain amplitude.}
    \label{fig:SPH_fourier_1}
\end{figure}

Results for SPH simulations of various values of the strain amplitude $\gamma_0$ as a function of $\omega/\omega_1$, with $\omega_1$ the frequency of the first peak, are shown in Figure \ref{fig:SPH_fourier_1}, normalised over the value of the first harmonic and obtained also thanks to the software MITLaos from \cite{MITLaos}. As expected, peaks at higher (odd) harmonics appear only for large displacements, and their relative intensity increases with the strain amplitude. Given the values of the Chebyshev coefficients shown in Figure \ref{fig:chebyshev_coefficients}, we can infer that this material exhibits intra-cycle strain stiffening ($e_3>0$) and the data suggests small shear thinning ($v_3 \lesssim 0$) at large displacements. This particular comparison is shown only over simulated data, as the experimental setup does not currently allow for the clean acquisition of data at very high strain amplitudes ($\gamma_0\ge10$).

\subsection{Comparison with experimental data}

\begin{figure}[tbp]
    \centering
    \begin{subfigure}{0.67\textwidth}
        \centering
        \includegraphics[width=\textwidth]{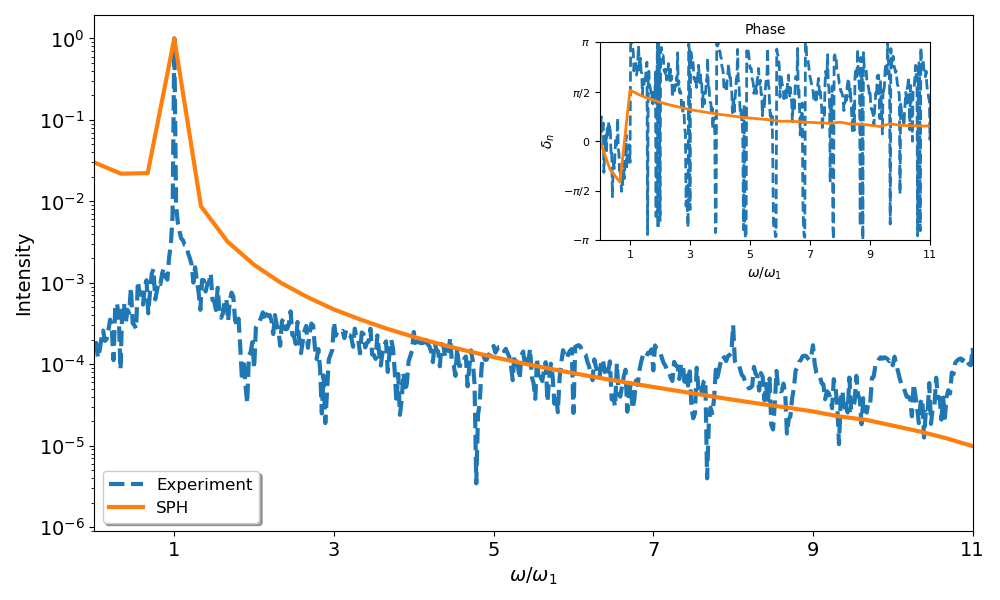}
        \caption{$\gamma_0=0.01$}
        \label{fig:small_displacement_comparison}
    \end{subfigure}
    \begin{subfigure}{0.67\textwidth}
        \centering
        \includegraphics[width=\textwidth]{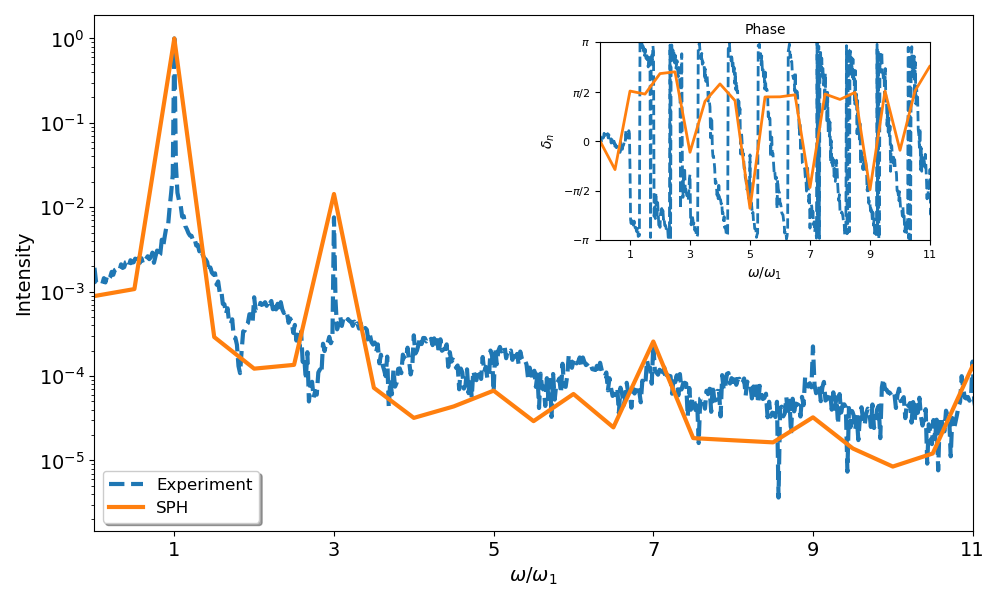}
        \caption{$\gamma_0=1.0$}
        \label{fig:medium_displacement_comparison}
    \end{subfigure}
    \begin{subfigure}{0.67\textwidth}
        \centering
        \includegraphics[width=\textwidth]{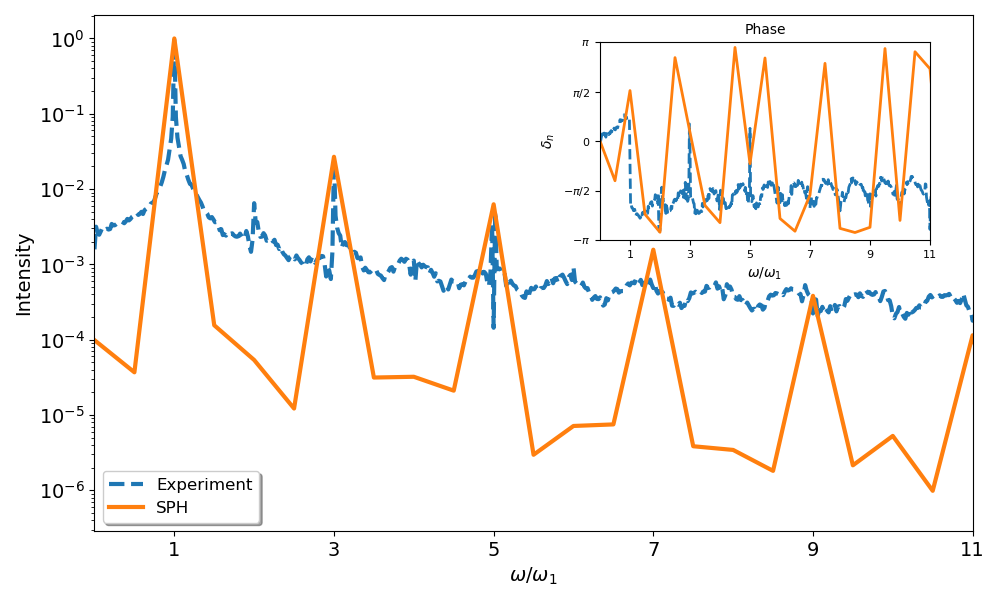}
        \caption{$\gamma_0=5.0$}
        \label{fig:large_displacement_comparison}
    \end{subfigure}
    \caption{Comparison of the FFT from experimental data and SPH simulation at $De\approx0.2$ (respectively \textcolor{cyan}{\rule[0.5ex]{0.4em}{0.8pt}~\rule[0.5ex]{0.4em}{0.8pt}} and \textcolor{orange}{\rule[0.5ex]{0.8em}{0.8pt}}). (a): small displacement; (b): medium displacement, (c): large displacement. Peaks at higher harmonics appear only for large displacements and their relative intensity matches the experimental data. In the inset, the phase $\delta$ is shown.}
    \label{fig:fourier_single}
\end{figure}

It can be shown that the the storage and loss moduli can be reconstructed from the Fourier transform of the stress signal by adding together the amplitude of the odd harmonics of the Fourier transform. In particular, the first harmonic reproduces the linear response of the material, and the ratio between the peaks of the first and third harmonics ($I_3/I_1$) is a good measure of the non-linearity of the material.  We compared the results of the SPH simulations with the experimental data, and we found a good agreement between the two regarding the appearance of non linearity (see Figure \ref{fig:fourier_single}), as well as the value of the ratio $I_3/I_1$ over the range of displacements explored. In particular, we found that the value $\gamma_0$ for which $I_3/I_1\sim1\%$, i.e. when the non-linearity becomes non negligible, is consistently around $1$. 
We wish to emphasise that, especially for large displacements, discrepancies between the experimental and numerical data can be observed at higher harmonics, which are more sensitive to noise. Indeed, we must consider that the experimental process is generally subject to small errors, including in the measurement of the stress signal, which can be amplified in the Fourier transform.
At the same time, the simulations employ a Lagrangian method involving moving interpolation points, which can introduce small errors in the computation of the stress. These numerical errors, while not impacting the main dynamic, produce small background noise for higher harmonics. Nevertheless, these discrepancies only appear at normalised intensities below $10^{-4}$, and do not affect the overall agreement between the experimental data and the numerical results.

From the values of the peaks, multiple non-equivalent definitions for the non-linear quantities of $G'$ and $G''$ can be used. \cite{ewoldt2008new} proposed two local definitions for the elastic modulus, the minimum-strain modulus $G'_M$ and the large-strain modulus $G'_L$, and for the dynamic viscosity: minimum rate viscosity $\eta_M$ and large-rate dynamic viscosity $\eta_L$. These values can be computed from the Fourier coefficients using the following equations:

\begin{align}
G'_M   &= \sum_{n\,\text{odd}} n G'_n
  &\qquad
G'_L   &= \sum_{n\,\text{odd}} (-1)^{(n-1)/2} G'_n \\
\eta_M &= \sum_{n\,\text{odd}} \frac{G''_n}{\omega}
  &\qquad
\eta_L &= \sum_{n\,\text{odd}} (-1)^{(n-1)/2} \frac{G''_n}{\omega}
\end{align}
where $G'_n$ and $G''_n$ are the Fourier coefficients of the stress signal. It is clear that for small displacements, where the additional harmonics are negligible, the sums are equivalent to their first term and therefore $G'_M = G'_L$. However, when non-linearity appears, differences between the two definitions can be observed. To quantify this effect and include additional informations about elastic and viscous behavious, the quantities $S=1-G'_M/G'_L$ and $e_3/e_1$, $v_3/v_1$ are often used \citep{ewoldt2008new} in addition to $G'_M$ and $G'_L$. We note that the ratios of Chebyshev coefficients conserve the information about the sign of the third coefficient, since $e_1$ and $v_1$ are always positive.

\begin{figure}[tbp]
    \centering
    \begin{subfigure}{0.48\textwidth}
        \centering
        \includegraphics[width=\textwidth]{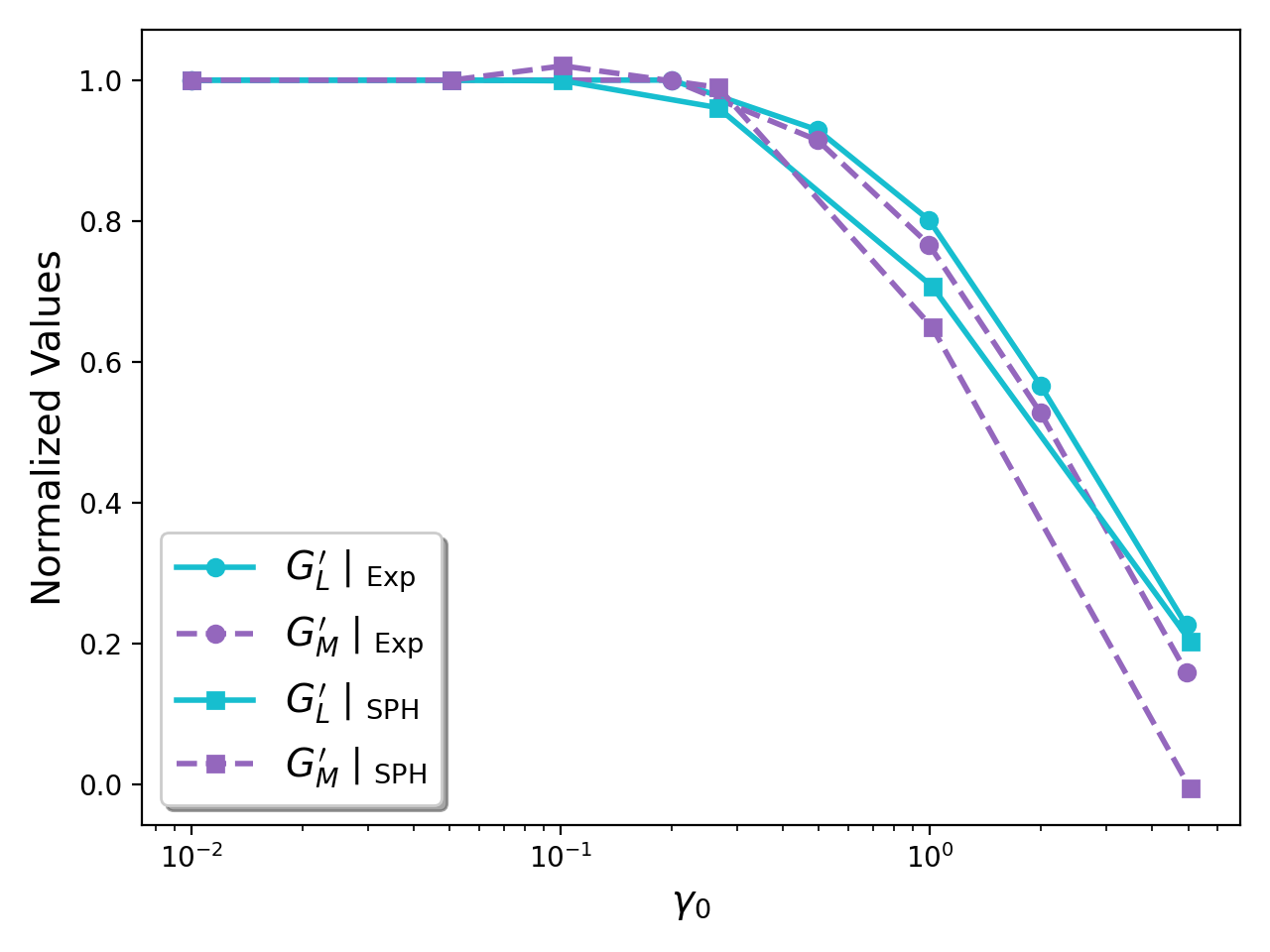}
        \caption{Minimum- and large-strain moduli comparison: SPH vs experiments.}
        \label{fig:GL_GM_SPH_vs_exp}
    \end{subfigure}
    \hfill
    \begin{subfigure}{0.48\textwidth}
        \centering
        \includegraphics[width=\textwidth]{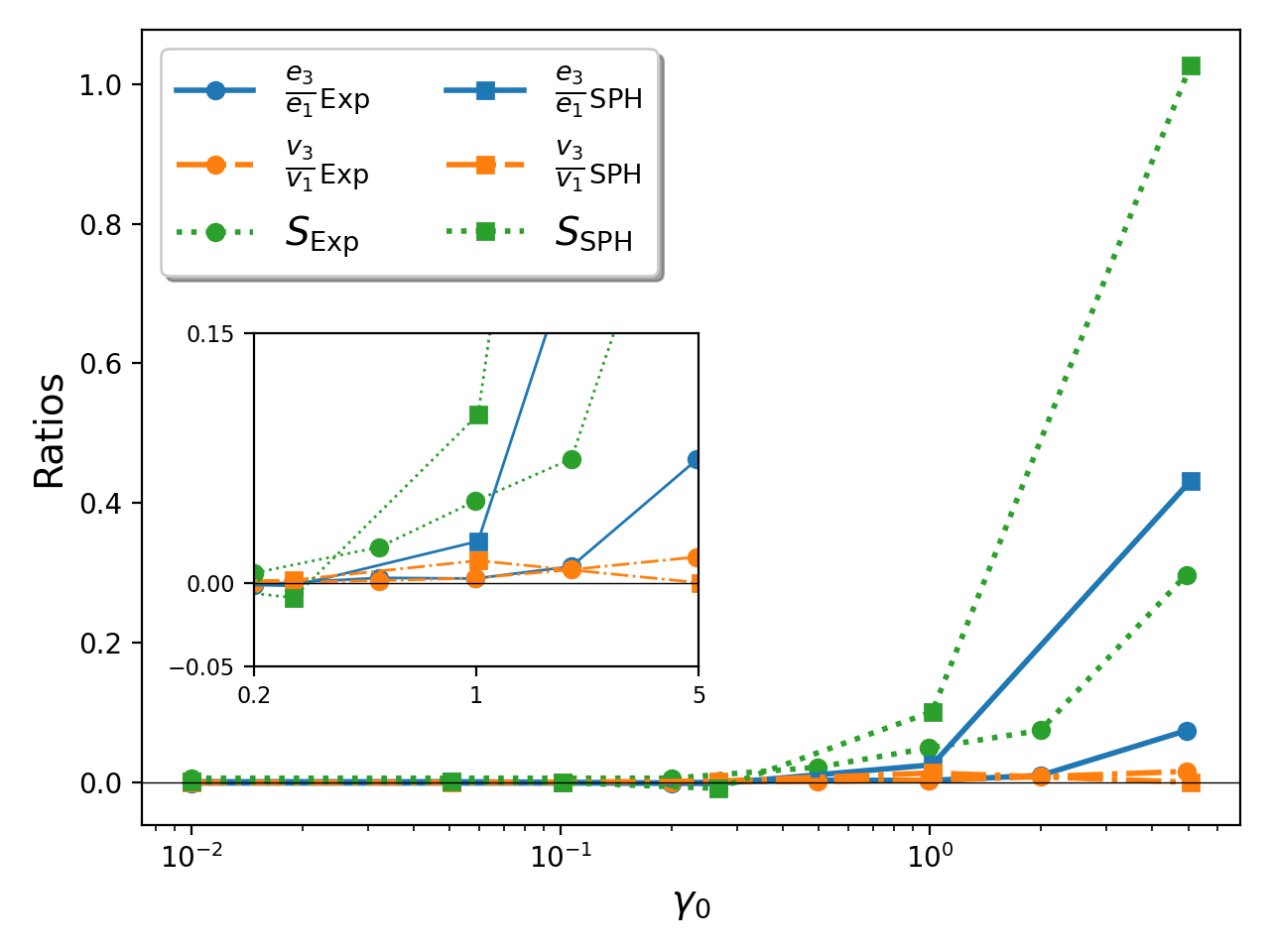}
        \caption{Behaviour of various ratios: SPH vs experiments.}
        \label{fig:S_e3_v3_SPH_vs_exp}
    \end{subfigure}
    \caption{(a) Comparison of the minimum-strain modulus $G'_M$  (\textcolor{purple}{- -}), large-strain modulus $G'_L$ (\textcolor{blue}{--}) as a function of strain amplitude $\gamma_0$ from experimental data ({\large$\bullet$}) and SPH simulations ({$\blacksquare$}) at $De\approx0.2$. The divergence between $G'_M$ and $G'_L$ at large amplitudes is a clear indicator of nonlinear viscoelastic behaviour. (b) Comparison of $e_3/e_1$ (\textcolor{blue}{--}) , $v_3/v_1$  (\textcolor{orange}{- -}) and $S=1-G'_M/G'_L$ (\textcolor{green}{:}) for experiments and SPH simulations at $De\approx0.2$.}
    \label{fig:GL_GM_and_var_all}
\end{figure}

Our analysis of the comparison between experimental data and SPH simulations is reported in Figure \ref{fig:GL_GM_and_var_all}. In particular, in Figure \ref{fig:GL_GM_SPH_vs_exp} we have computed the values of $G'_M$ and $G'_L$ from both experimental and numerical data at $De\approx0.2$, and we observe a general agreement between the two lines, showing a non-linearity growing with $\gamma_0$. However, we can also note that the non-linear behaviour of the material seems to grow slightly faster in the numerical simulations compared to the experimental data, which suggests that the damping function $h(\gamma)$ could be further optimised. A further analysis on the effect of the non-linearity is shown in Appendix \ref{app:damping_effects}.
Figure \ref{fig:S_e3_v3_SPH_vs_exp} shows the comparison between experimental and numerical data for the quantities $S$, $e_3/e_1$ and $v_3/v_1$ at $De\approx0.2$. Also in this case, we observe a general agreement between the two sets of data, in particular regarding the sign of the ratios, with the non-linear effects becoming more pronounced at larger values of $\gamma_0$. As before, we note that the non-linear behaviour seems to grow faster in the numerical simulations compared to the experimental data. 

\begin{figure}[tbp]
    \centering
    \begin{subfigure}{0.48\textwidth}
        \centering
        \includegraphics[width=\textwidth]{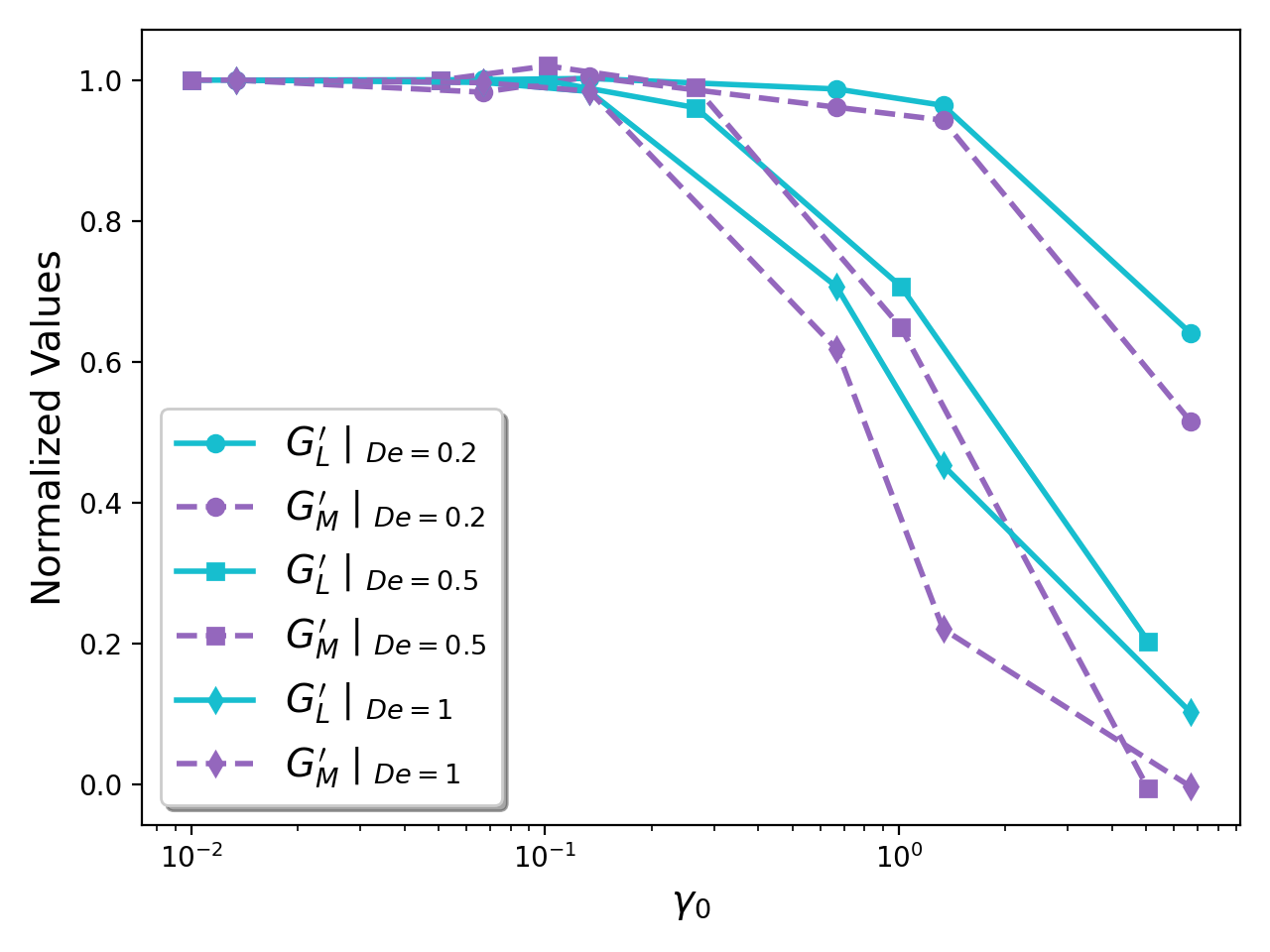}
        \caption{Minimum- and large-strain moduli comparison for different $De$ as a function of $\gamma_0$.}
        \label{fig:GL_GM_De}
    \end{subfigure}
    \hfill
    \begin{subfigure}{0.48\textwidth}
        \centering
        \includegraphics[width=\textwidth]{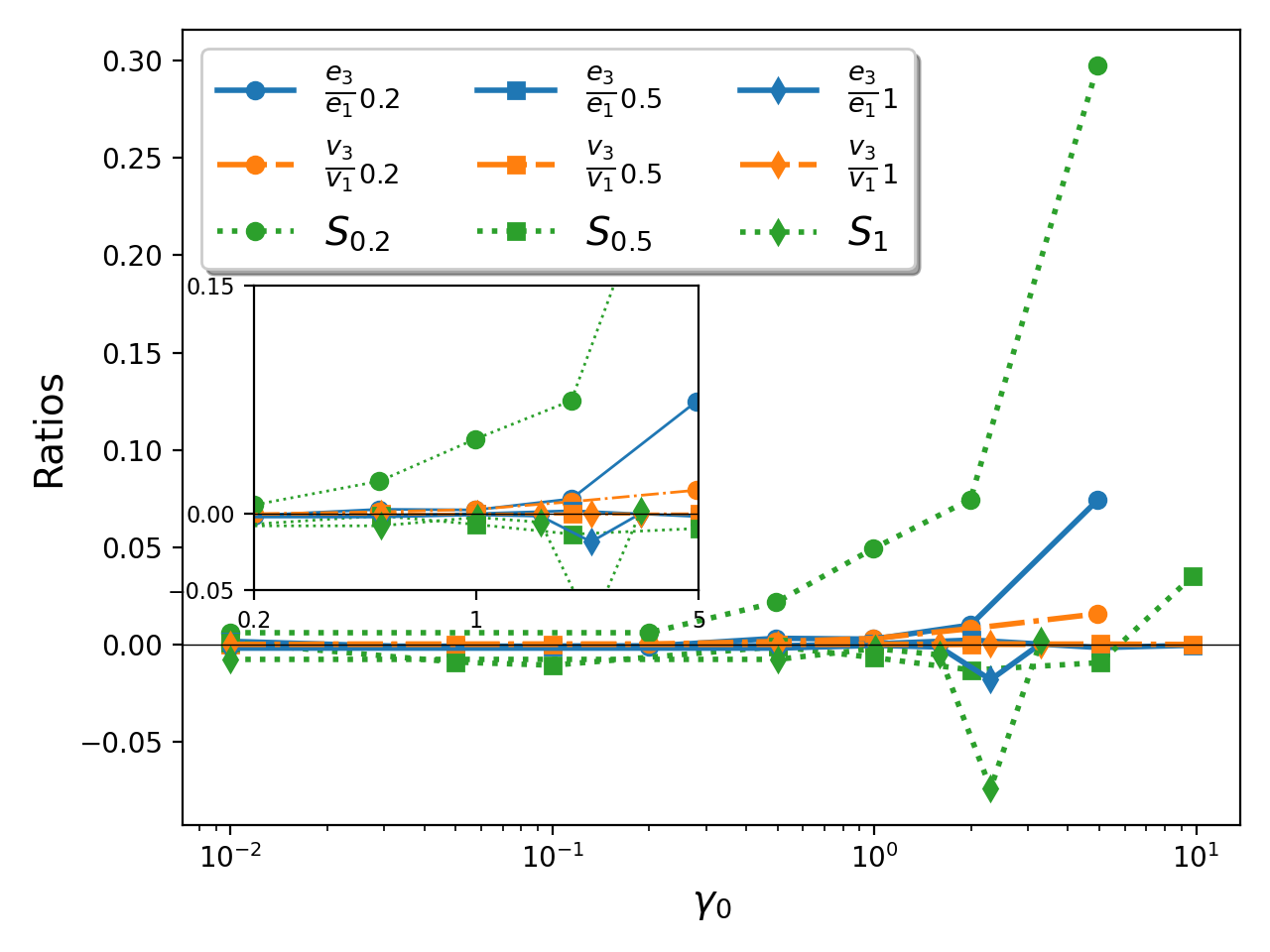}
        \caption{Behaviour of various ratios for different $De$ as a function of $\gamma_0$.}
        \label{fig:S_e3_v3_De}
    \end{subfigure}
    \caption{(a) Comparison of the minimum-strain modulus $G'_M$  (\textcolor{purple}{- -}), large-strain modulus $G'_L$ (\textcolor{blue}{--}) as a function of strain amplitude $\gamma_0$ from experimental data at $0.2\lesssim De\lesssim1$. (b) Comparison of $e_3/e_1$ (\textcolor{blue}{--}) , $v_3/v_1$  (\textcolor{orange}{- -}) and $S=1-G'_M/G'_L$ (\textcolor{green}{:}) for experiments at $0.2\lesssim De\lesssim1$.}
    \label{fig:GL_GM_and_var_all_De}
\end{figure}

A comparison of data from  experiments at different Deborah number has also been performed, and the results are shown in Figure \ref{fig:GL_GM_and_var_all_De}. We observe that generally the non-linear effects for large displacements are more pronounced at larger values of $De$, as expected. On the other hand, the onset of non-linearity for medium values of $\gamma_0$ seems to occur earlier for lower values of the oscillation frequency. However, we note that the differences between the various $De$ values are not very pronounced, suggesting that the non-linear behaviour is mostly governed by the strain amplitude $\gamma_0$ rather than the frequency of oscillation $\omega_0$ within the range of $De$ explored.

\section{Conclusion}\label{sec:conclusions}

In this work, we have presented a numerical implementation of the fractional K-BKZ model for the simulation of the rheological behaviour of polypropylene-based polymer melts under large amplitude oscillatory shear (LAOS) using the Smoothed Particle Hydrodynamics (SPH) method. The SPH method is a Lagrangian particle-based approach that allows for the simulation of complex fluid flows with large deformations and complex geometries. We have shown that the SPH method is well-suited for implementing integral models, such as the fractional K-BKZ model, which captures the nonlinear viscoelastic behaviour of polymer melts.

Indeed, once the model parameters are fitted to the experimental values in the linear regime, and a damping function is identified, the SPH simulations recover the occurrence of non-linearity in a consistent way when compared to new experimental data for all the strain amplitudes tested. Therefore, we believe that this implementation has the potential for being used in a wide range of applications, to predict the experimental behaviour of polymer melts under LAOS conditions, and to provide insights into the underlying mechanisms of viscoelasticity in complex fluids, helping the technological development.

The analysis of the system under more complex geometrical conditions is a natural next step. We are currently investigating the case for which the source of non linearity is the geometry, rather than the imposed flow, i.e. the case of a polymer melt flowing in a channel with a sudden expansion or contraction. In this case, the flow is not uniform and the shear rate is not constant. Finally, a topic for a future study will be the analysis of a viscous matrix filled with solid particles, which is a common scenario in many industrial applications, such as the processing of polymer composites or the production of polymer-based materials with specific mechanical properties. This would extend the initial promising SPH results presented in \cite{vazquezquesada2019shear} for a viscoelastic suspension with a simple Oldroyd-B matrix, which also motivated the choice of SPH-implemented fractional derivatives for this study, rather than adapting existing kernels for fractional derivatives or relying on sparse polynomial interpolations \citep{podlubny2009matrix,shanbhag2024sparse}.
\backmatter





\bmhead{Acknowledgements}

This research is supported by the Basque Government through the BERC 2022-2025 program, the ELKARTEK 2024 program (ELASTBAT: KK- 2024/00091).  The research is also partially funded by the Spanish State Research Agency through BCAM Severo Ochoa excellence accreditation CEX2021-0011 42-S/MICIN/AEI/10.13039/501100011033, and through the project PID2024-158994OB-C42 (‘Multiscale Modeling of Friction, Lubrication, and Viscoelasticity in Particle Suspensions’ and acronym ‘MMFLVPS’) funded by MICIU/AEI/10.13039/501100011033 and cofunded by the European Union.










\begin{figure}[tbp]
    \centering
    \begin{subfigure}{0.43\textwidth}
        \centering
        \includegraphics[width=\textwidth]{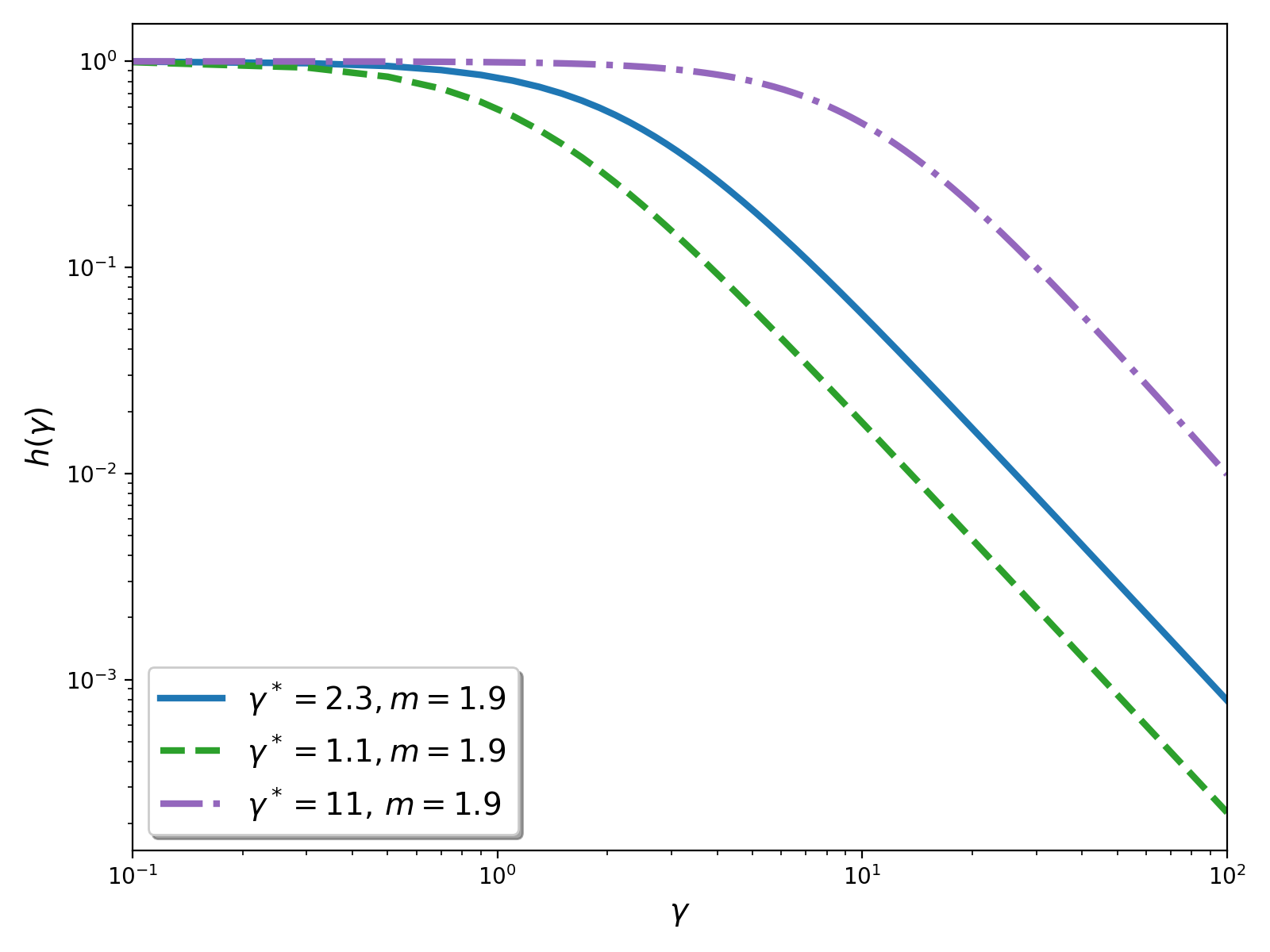}
        \caption{Damping functions for different values of the parameters $a$ and $b$.}
        \label{fig:damping_functions}
        \end{subfigure}
    \hfill
    \begin{subfigure}{0.55\textwidth}
        \centering
        \includegraphics[width=\textwidth]{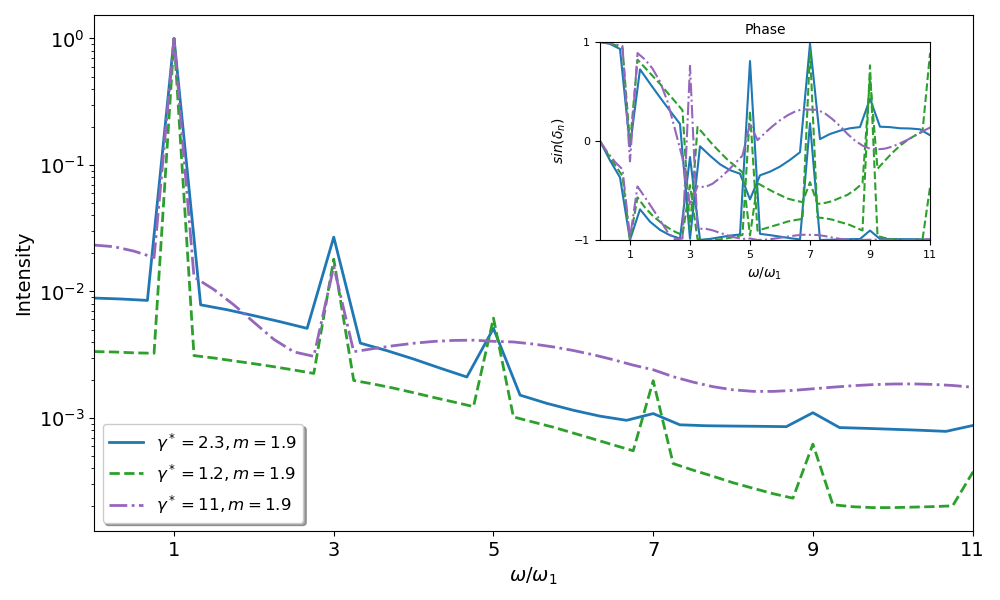}
        \caption{Comparison of the Fourier coefficients for different values of the damping function parameters.}
        \label{fig:damping_effects}
    \end{subfigure}
    \caption{(a) Damping functions $h(\gamma)$ for different values of the parameters. In blue, the curve fitting the experimental data. (b) Comparison of the Fourier coefficients for different values of the damping function parameters at $\gamma_0=5$ and $De=0.2$. Stronger damping leads to higher relative intensities at the higher harmonics.} 
    \label{fig:damping_analysis}
\end{figure}

\begin{figure}[bpt]
    \centering
    \begin{subfigure}{0.48\textwidth}
        \centering
        \includegraphics[width=\textwidth]{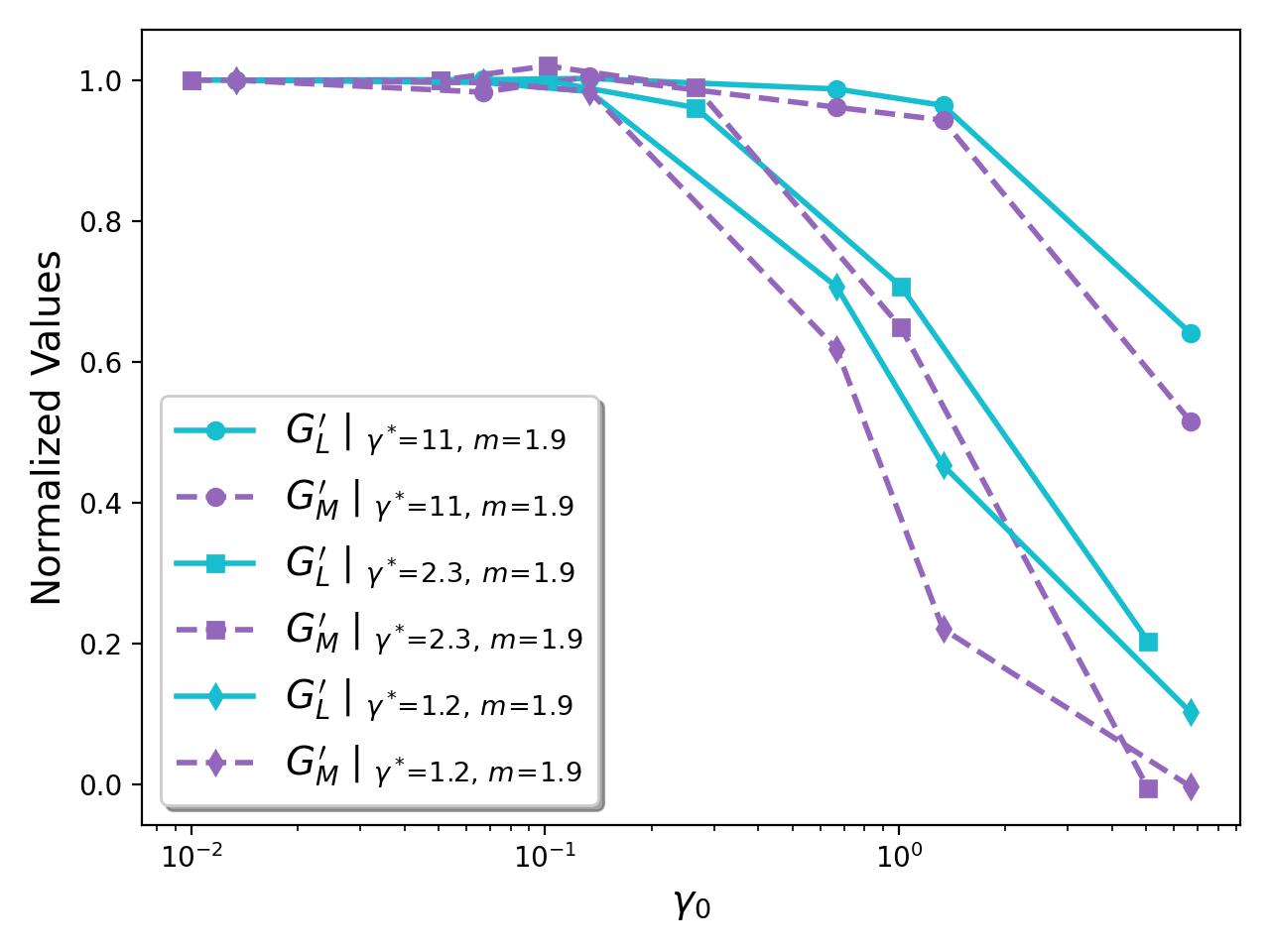}
        \caption{Minimum- and large-strain moduli comparison for different damping functions.}
        \label{fig:GL_GM_damping_effects}
    \end{subfigure}
    \hfill
    \begin{subfigure}{0.48\textwidth}
        \centering
        \includegraphics[width=\textwidth]{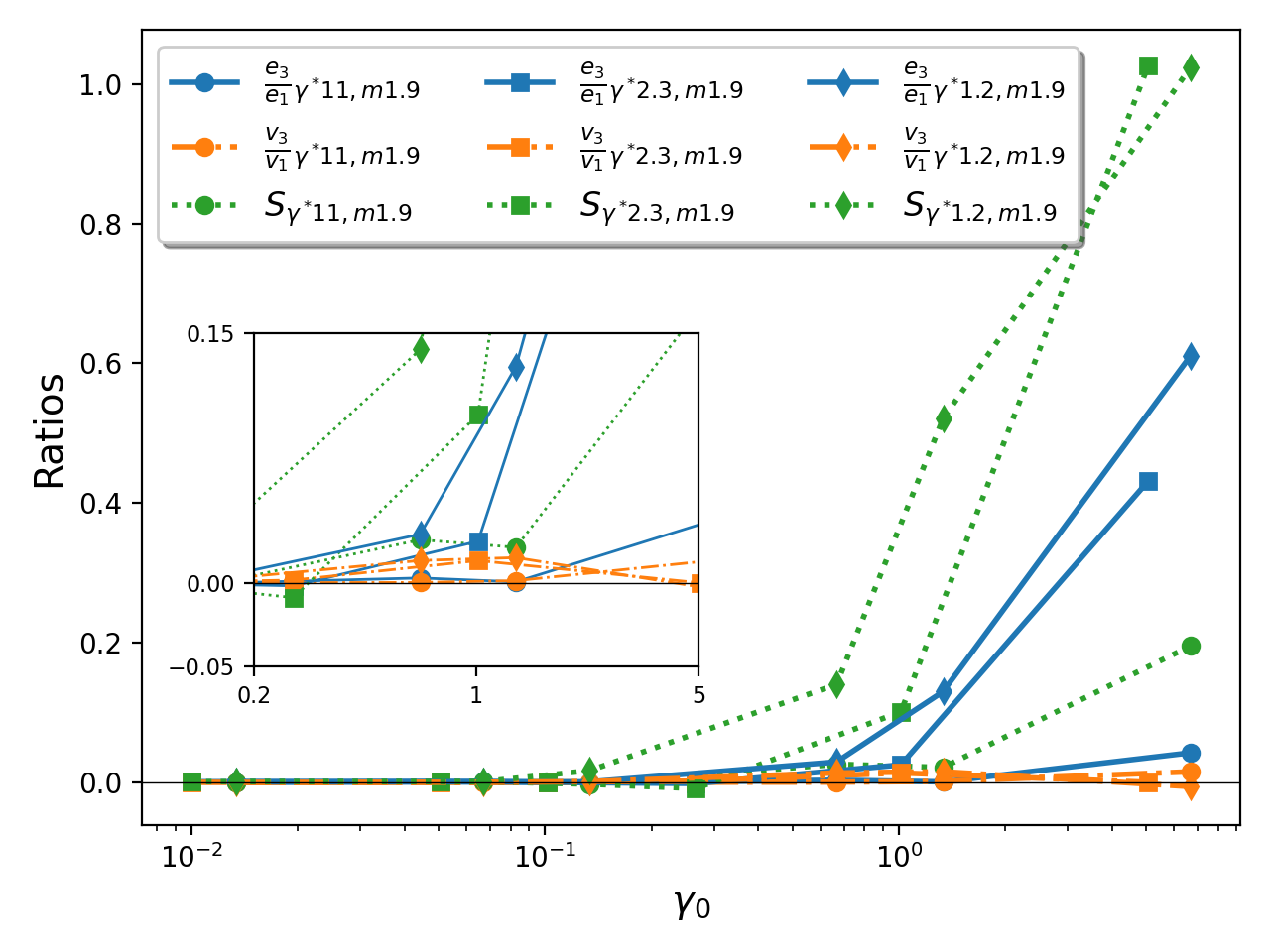}
        \caption{Behaviour of various ratios for different damping functions.}
        \label{fig:S_e3_v3_damping_effects}
    \end{subfigure}
    \caption{(a) Comparison of the minimum-strain modulus $G'_M$  (\textcolor{purple}{- -}), large-strain modulus $G'_L$ (\textcolor{blue}{--}) as a function of strain amplitude $\gamma_0$ from SPH simulations with different damping function parameters at $De\approx0.2$. The divergence between $G'_M$ and $G'_L$ at large amplitudes is a clear indicator of nonlinear viscoelastic behaviour. (b) Comparison of $e_3/e_1$ (\textcolor{blue}{--}) , $v_3/v_1$  (\textcolor{orange}{- -}) and $S=1-G'_M/G'_L$ (\textcolor{green}{:}) for SPH simulations with different damping function parameters at $De\approx0.2$.} 
    \label{fig:GL_GM_and_var_all_damping}
\end{figure}

\begin{appendices}
\section{Effect of the damping function parameters}\label{app:damping_effects}

In this appendix, we present a brief analysis of the effects of the parameters of the damping function $h(\gamma)$ on the results of the SPH simulations. The damping function is defined in equation \eqref{eq:damping_function} as $h(\gamma) = (1+(\gamma/\gamma^*)^m)^{-1}$, where $\gamma^*$ is the critical strain parameter and $m$ is the exponent that controls the rate of strain thinning.

We have performed a series of SPH simulations with varying the parameters of the damping function, while keeping the other parameters of the fractional K-BKZ model constant. The chosen damping functions are shown in Figure \ref{fig:damping_functions}, where it is shown that, as expected, the onset of non linearity is proportional to $1/\gamma^*$. To analyse the effect of the changes in damping function, we have then analysed the Fourier transform for each simulation. In Figure \ref{fig:damping_effects}, we show the results of the Fourier transform for $\gamma_0=5$ and $De=0.2$. It can be observed that the choice of the parameters of the damping function has a significant effect on the relative intensity of the higher harmonics, with lower values of $\gamma^*$ leading to higher intensities at the higher harmonics. This is consistent with the expectation that a stronger damping leads to a more pronounced non-linear behaviour.

In Figure \ref{fig:GL_GM_damping_effects}, we analysed how the choice of the damping function parameters affects the values of $G'_M$ and $G'_L$. We can observe that stronger damping (lower $\gamma^*$) leads to an earlier and more pronounced divergence between $G'_M$ and $G'_L$ at a given $\gamma_0$, indicating a stronger non-linear behaviour. A similar behaviour can be observed in Figure \ref{fig:S_e3_v3_damping_effects}, where stronger damping leads to an increase in $S$ and the Chebyshev ratios for the same imposed shear.

While the hypothesis of time-strain separability that underpins the K-BKZ model may not be strictly valid for all materials, this analysis suggests that a careful selection of the damping function parameters can help to capture the non-linear viscoelastic behaviour of polymer melts under LAOS conditions, particularly concerning the effects on $e_3$ and $v_3$. Further experimental data and analysis would be needed to fully validate this approach for specific materials different from the one considered here.




\end{appendices}

\pagebreak


\end{document}